\documentclass[sigconf]{acmart}

\usepackage{enumitem}
\usepackage{graphicx}
\usepackage{footnote}
\usepackage{footmisc}

\AtBeginDocument{%
  }

\acmPrice{15.00}
\acmISBN{978-1-4503-XXXX-X/18/06}

\renewcommand{\thefootnote}{\alph{footnote}}

\newcommand{\astfootnote}[1]{%
\let\oldthefootnote=\thefootnote%
\setcounter{footnote}{0}%
\renewcommand{\thefootnote}{\fnsymbol{footnote}}%
\footnote{#1}%
\let\thefootnote=\oldthefootnote%
}

\begin{document}

\title{Deepfake CAPTCHA: A Method for Preventing Fake Calls}



\author{Lior Yasur,$^*$ Guy Frankovits,$^*$ Fred M. Grabovski, Yisroel Mirsky}%
 \email{{lioryasu,guyfrank,freddie}@post.bgu.ac.il,          yisroel@bgu.ac.il}%





 \affiliation{%
   \institution{Ben-Gurion University of the Negev}
    \country{Israel}
  }
\renewcommand{\shortauthors}{Yasur et al.}

\begin{abstract}
\footnotetext{$^*$These authors have equal contribution}

Deep learning technology has made it possible to generate realistic content of specific individuals. These `deepfakes' can now be generated in real-time which enables attackers to impersonate people over audio and video calls. Moreover, some methods only need a few images or seconds of audio to steal an identity. Existing defenses perform \textit{passive} analysis to detect fake content. However, with the rapid progress of deepfake quality, this may be a losing game.

In this paper, we propose D-CAPTCHA: an \textit{active} defense against real-time deepfakes. The approach is to force the adversary into the spotlight by challenging the deepfake model to generate content which exceeds its capabilities. By doing so, passive detection becomes easier since the content will be distorted. In contrast to existing CAPTCHAs, we challenge the AI's ability to create content as opposed to its ability to classify content. In this work we focus on real-time audio deepfakes and present preliminary results on video.  

In our evaluation we found that D-CAPTCHA outperforms state-of-the-art audio deepfake detectors with an accuracy of 91-100\% depending on the challenge (compared to 71\% without challenges). We also performed a study on 41 volunteers to understand how threatening current real-time deepfake attacks are. We found that the majority of the volunteers could not tell the difference between real and fake audio.

\end{abstract}




\maketitle

\section{Introduction}

\begin{figure}
    \centering
    \includegraphics[width=0.7\columnwidth]{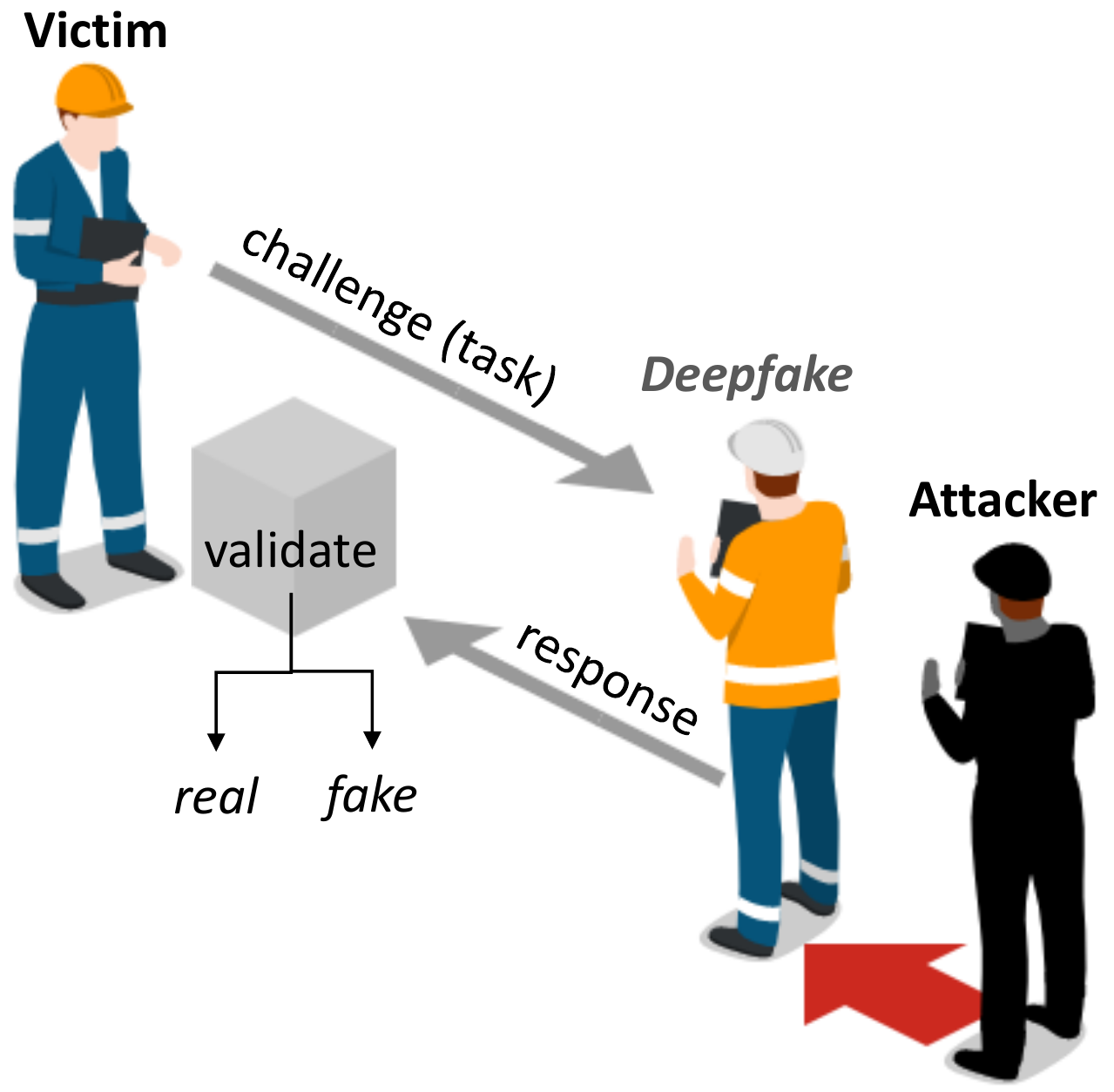}
    \caption{Overview of the proposed defense: the victim requests the caller to perform a task which is challenging for a deepfake model to perform. If the response is distorted or does not contain the task, then the caller is likely a deepfake.}
    \label{fig:overview}
\end{figure}


A deepfake is any media, generated by a deep neural network, which is authentic from a human being's perspective \cite{mirsky2021creation}. Since the emergence of deepfakes in 2017, the technology has improved in terms of quality and has been adopted in a variety of applications. For example, deepfake technology is used to  enhance productivity \cite{VoiceClo10:online}, education \cite{Deepfake15:online} and provide entertainment \cite{ShamookS49:online}. However, the same technology has been used for unethical and malicious purposes as well. For example, with a deepfake, anyone can impersonate a target identity by reenacting the target's face and/or voice. This ability has enabled threat actors to perform defamation, blackmail, misinformation, and social engineering attacks on companies and individuals around the world \cite{Reshapin92:online}. 
For example, since 2017, the technology has been used to `swap' the identity of individuals into explicit videos for unethical \cite{Deepfake76:online} and malicious \cite{Cheerlea79:online} reasons. More recently, in March 2022 during the Russian-Ukraine conflict, a deepfake video was circulated depicting the prime minister of Ukraine telling his troops to give up and stop fighting \cite{Deepfake46:online}. 

\subsection{Real-time Deepfakes (RT-DF)}
Deepfake technology has improved over the last few years in terms of efficiency. This has enabled attackers to create real-time deepfakes (RT-DF)\footnote{Examples of RT-DF tools:
\url{https://github.com/iperov/DeepFaceLive}\\
\url{https://github.com/alievk/avatarify-python}\\
\url{https://samsunglabs.github.io/MegaPortraits/}\\
\url{https://www.respeecher.com/}}\\
With an RT-DF, an attacker can impersonate people over voice and video calls. The danger of this emerging threat is that (1) the attack vector is not expected, (2) familiarity can be mistaken as authenticity and (3) the quality of RT-DFs is constantly improving.

To conceptualize this threat, let's perform the following thought experiment. Imagine someone receives a call from their mother who is in trouble and urgently needs a money transfer. The caller sounds exactly like her, but the situation seems a bit out of place. Under stress and frustration, she hands the phone over to someone who sounds like the victim's father, who confirms the situation. Without hesitation, many would transfer the money even though they're technically talking to a stranger.
Now consider state-actors with considerable amounts of time and resources. They could target workers at power plants and other critical infrastructure by posing as their administrators. Over a phone call, they could convince the worker to change a configuration or reveal confidential information which would lead to a cyber breach or a catastrophic failure. Attackers could even pose as military officials or politicians leading to a breach of national security. 

These scenarios are plausible because some existing real-time frameworks can impersonate an individual's face or voice using very little information. For example, some real-time methods can reenact a face with one sample image \cite{drobyshev2022megaportraits,NIPS2019_8935} and some can clone a voice with just a few seconds of audio \cite{AdaIN,FragmentVC}. Using these technologies, an attacker would only need to call the source voice for a few seconds or scrape the source's image from the internet to perform the attack.

\subsection{The Emerging Threat of RT-DFs}
Threat actors already understand the utility of RT-DFs. This is evident in recent events where RT-DFs have been used to perform criminal acts. The first case was discovered in 2019 when a CEO was tricked into transferring \$243k due to an RT-DF phone call \cite{Fraudste87:online}. In 2021, senior European MPs participated in Zoom meetings with someone masquerading as Russian opposition figures \cite{European87:online}. In the same year, cyber criminals pulled off a \$35 million bank heist involving RT-DF audio calls to a company director, tricking him to perform money transfers \cite{Fraudste98:online}. In June 2022, the FBI released a warning that cyber criminals are using RT-DFs in job interviews in order to secure remote work positions and gain insider information. Then in August that year, cyber criminals attended Zoom meetings masquerading as the CEO of Binance \cite{Binancee98:online}.

\subsection{The Gap in Current Defenses}
Many methods have been proposed for detecting deepfakes \cite{mirsky2021creation,almutairi2022review}. These methods typically use deep learning models to either (1) detect mistakes or artifacts in generated media, or (2) search for forensic evidence such as a latent noise patterns (examples of these works can be found in section \ref{sec:relwork}). However, there are two fundamental problems with existing defenses:
\begin{description}[leftmargin=.5cm]
    \item[Longevity.] Methods which identify semantic errors or artifacts have the assumption that the quality of deepfakes will not significantly improve. However, it is clearly evident that the quality of deepfakes \textit{is} improving  and at a fast rate \cite{masood2022deepfakes}. Therefore, artifact-based methods have a high potential of becoming obsolete within a short time-frame.
    \item[Evasion.] Methods which rely on latent noise patterns can be evaded by applying a post-processor. For example a deepfake can be passed through a low pass filter, undergo compression or be given additive noise. Moreover, these processes are common in audio and video calls. Therefore, the attacker may not need to do anything to remove the forensic evidence in the call.
\end{description}

\subsection{Real-Time CAPTCHA}
In this paper, we propose Deepfake-CAPTCHA (\textbf{D-CAPTCHA}): a system for automatically detecting deepfake calls through challenge response analysis. Instead of passively observing call content, we actively interact with the caller by requesting that he or she to perform a task (the \textit{challenge}). The task is easy for a human to perform but extremely hard for a deepfake model to recreate due to limitations in attack practicality and technology. When a deepfake tries to perform the task, the resulting content (the \textit{response}) will be severely distorted --making it easier for an anomaly detector, classifier, or even the victim to detect. In addition, we propose using an identity model and task detection model to mitigate evasion tactics. The identity model compares the identity of the caller before and during the response to ensure that the caller cannot turn off the RT-DF during the task or splice in content from other identities. Similarly, the task detection model ensures that the caller has indeed performed the task as opposed to doing nothing. 

Existing CAPTCHA systems, such as reCAPTHCA,\footnote{\url{https://developers.google.com/recaptcha/}} challenge AI to interpret content. In contrast, we propose a system which challenges AI to \textit{create content}, with additional constraints on realism, identity, task (complexity), and time. 

In this work, we focus on audio-based RT-DF attacks (voice cloning). We consider audio RT-DFs a more significant threat over video RT-DFs because it is easier for an attacker to make a phone call than setup a video call with the victim. Also, their occurrences in the wild are increasing \cite{AIgenera5:online}.  Therefore, RT-DF audio calls are arguably a bigger threat at this time. However, we note that the same D-CAPTCHA system proposed in this paper can be applied to video calls as well. In section \ref{sec:futurework} we present initial results in this domain.

In our evaluation, we collected five state-of-the-art audio RT-DF technologies. We performed a panel survey to see what the public thinks about their quality and we evaluated the top two models on our defense and on others as well. We found that our method can significantly enhance the performance of state-of-the-art audio-based deepfake detectors.

\subsection{Contributions}
In summary, our work has the following contributions:
\begin{itemize}
    \item We propose the first active defense against RT-DFs. Compared to existing artifact-base methods, our approach (1) provides stronger guarantees of detection than using only passive detection and (2) has better longevity because the challenges are extensible.
    \item We define what a D-CAPTCHA is and what constitutes a strong deepfake CAPTCHA: We identify the limitations of existing RT-DF systems and propose four constraints a challenge must present to a caller. We also present how these constraints can be verified in a response both manually and automatically. We also provide an initial set of CAPTCHAs and analyze their security and usability.
    \item We evaluated the quality of five state-of-the-art RT-DF voice cloning models with 41 volunteers. Doing so enables us to better understand the current threat which this technology poses.
    \item We provide thorough evaluations on (1) how well the CAPTCHA system performs and (2) how robust it is against an evasive adversary.

\end{itemize}
%

\section{Background}
In this work, we focus on mitigating the threat of real-time voice cloning. Furthermore, we focus on methods that perform speech-to-speech voice conversion (VC) \cite{Assem-VC,AutoVC-F0,FragmentVC,cyclegan-VC-V3,MediumVC,StarGANv2-VC} as opposed to text-to-speech (TTS) methods such as \cite{jia2018transfer}. 

Let $t$ be a target identity which we'd like to clone, and $a_s$ be an audio clip of identity $s$ speaking. Content is the part of speech that is independent of a speaker's vocal anatomy (e.g., words, accent, enunciation, and so on).
The objective of voice cloning is to perform $f_t(a_s)=a_g$ where $a_g$ is generated audio containing the content of $a_s$ in the style of $t$. In an attack, $t$ is an individual who is familiar to the victim, and $s$ is the attacker (or a voice actor hired by the attacker).

To convert unbounded audio streams in real-time, audio is processed as a sequence of short audio frames (approximately 10-1000ms each). In this way, the $i$-th input frame $a^{(i)}_t$ is converted into $a^{(i)}_g$ within one second. We consider $f_t$ to be an RT-DF if the pipeline can be executed with no more than a $1$ second delay from the microphone to speaker. In other words, the time it takes for an utterance to be recorded, converted, and played back is no longer than 1 second. Longer delays may raise the victim's suspicion. Methods which process entire recordings all at once form non-casual systems. Therefore, we do not consider them as RT-DF systems (e.g., \cite{rebryk2020convoice}). 

There are various levels of flexibility when it comes to prior knowledge of $s$ and $t$. For instance, not every model can drive $a_g$ with content from $s$ without prior training on $s$. Many of the audio RT-DF models can be categorized as follows:
\begin{description}[leftmargin=.5cm]
    \item[many-to-many.] Are models which require both the source voice $s$ (used in $c$) and the target voice $t$ to be in $f$'s training set. Since $s$ is the attacker, the only challenge is collecting samples of $t$. 
    \item[any-to-many.]  Are models which can use any source voice to drive the content in $x_g$ without retraining the model.
    \item[any-to-any.] Are models which do not need to see the source $s$ or target $t$ during training to perform $f_t(c_s)=x_g$. This makes any-to-any models the flexible solution for attackers.
\end{description}

\section{Threat Model}
There are two ways an adversary can use the RT-DF $f_t$ maliciously: the adversary can (1) call a victim while impersonating $t$ or (2) call a target and threaten to impersonate him. The call may take place over the phone through a virtual meeting (such as over Zoom). We refer to these calls as ``fake calls''.

\subsection{Attack Goals}
There are several attack goals which an adversary can achieve using a fake call:
\begin{description}[leftmargin=.5cm]
  \item[Cyber attacks.] Fake calls can be used in social engineering attacks (SE). For example, instead of sending spear phishing emails to get employees to install malware, the attacker can call a victims up as their manager and ask them to do it directly. These SE attacks can also be used during an adversary's reconnaissance on an organization to obtain system information and credentials. For example, the attacker can call a victim posing as a colleague, asking for help to login or claiming that he has "forgotten" some information.
  \item[Sabotage.] An attacker can impersonate a victim's supervisor in an attempt to have the victim change some settings or configurations in a system. For example, in a chemical processing plant, an adversary can use a manager's voice to tell a worker to urgently alter the balance of some process --leading to catastrophic results.
  \item[Espionage.] Fake calls can also be used by state agents as a means for extracting sensitive and confidential information. For example, an adversary can gain a political advantage by posing as a politician's assistant and a military advantage by posing as a military official. Moreover, sensitive documents and source code can be leaked in a similar manner if the adversary impersonates a leading figure who directly asks employees for this material. Finally, by impersonating professionals with LinkedIn profiles, an adversary can obtain remote job interviews which may lead to remote work with a company --ultimately placing an insider within the organization \cite{Internet56:online}.
  \item[Scams.] An attacker can prey upon people and trick them into giving them money. For example, the adversary can impersonate a family member of the victim to convince the victim that his family is in danger and needs an urgent money transfer. Similar schemes can be done on business and banks where the attacker convinces the victim to make a money transfer under false pretexts \cite{Fraudste87:online,Fraudste98:online}.
  \item[Blackmail.] To coerce a victim to perform an action (pay money, reveal information, ...) an attacker can blackmail the victim using RT-DF technology. For example, the attacker can speak to the victim using the victim's voice and threaten the victim that calls will be made to reporters, friends, colleagues, or a spouse as the victim if the blackmail terms are not met (similar to a case that happened in Singapore \cite{Singapor7:online}).
  \item[Defamation.] An adversary can defame the victim by performing embarrassing or unethical acts over calls to the victim's colleagues or reporters while masquerading as the victim.
  \item[Misinformation.] An attacker can call reporters and do interviews as politicians and other public figures to spread misinformation in the media.
\end{description}

\subsection{Attack Setup}
The flexibility of the attacker depends on the flexibility of the RT-DF model. To train the model $f_t$, the attacker can use one of two common approaches:
\begin{description}[leftmargin=.5cm]
    \item[Batch Learning.] If the attacker uses conventional learning models such as 
    \cite{StarGANv2-VC,Assem-VC,cyclegan-VC-V3,AutoVC-F0}, then the attacker will need to collect a large audio training set of $t$ (typically around 20-30 minutes) and train $f$ on this data. This dataset can be obtained from the Internet if $t$ is a celebrity (e.g., interviews on YouTube). If $t$ doesn't have an internet presence, then the dataset may be obtained via long phone calls, wiretaps, and secret recordings (bugs). These models are usually \textbf{many-to-many} or \textbf{any-to-many}.

    \item[Few/Zero-shot Learning.] When using methods such as \cite{AdaIN,FragmentVC,MediumVC,zeroshot}, the attacker only needs a few seconds of $t$'s audio. In this case, the attacker can make a short phone call to $t$ and record his/her voice. The adversary may also find short video clips on social media or resort to wiretaps and bugs as well.  These types of models are usually \textbf{any-to-any}.
    
\end{description}
We note that most modern RT-DF technologies do not require labeled data since they are trained in a self-supervised manner \cite{mirsky2021creation}. Regarding quality, batch model training methods are typically preferred over few-shot or zero-shot methods.

\section{Related works}\label{sec:relwork}




Most audio deepfake detection systems (ADDS) use a common pipeline to detect deepfake audio: given an audio clip $a$, the pipeline (1) converts $a$ into a stream of one or more audio frames $a^{(1)},... a^{(n)}$, (2) extracts a feature representation from each frame which summarizes the frames' waveforms $x^{(1)},... x^{(n)}$, and then (3) passes the frame(s) through a detector which predicts the likelihood of $a$ being real or fake. The audio features in $x^{(i)}$ are either a Short Time Fourier Transform (STFT) \cite{alzantot2019deep, wijethunga2020deepfake}, spectrogram, Mel Frequency Cepstral Coefficients (MFCC) \cite{subramani2020learning, khalid2021evaluation}, or the Constant Q Cepstral Coefficients (CQCC) \cite{lei2020siamese, lai2019assert} of $a^{(i)}$. Some methods simply use the actual waveform of $a^{(i)}$  \cite{tak2021end, rawnet2}. 

With this representation, an ADDS can either use a classifier \cite{kawa2022specrnet, rawnet2, khochare2022deep} or anomaly detector \cite{khalid2020oc, alegre2013one} to identify generated audio. A good summary of modern ADDS can be found in \cite{almutairi2022review}. In general, classifiers are trained on labeled audio data consisting of two classes: real and deepfake. By providing labeled data, the model can automatically identify the relevant features (semantic or latent) during training. An intuitive example is the case where a deepfake voice cannot accurately pronounce the letter `B' \cite{agarwal2020detecting}. In this scenario, the model will consider this pattern as a distinguishing feature for that deepfake. A disadvantage of classifiers is that they follow a closed-world assumption; that all examples of the deepfake class are in the training set. This assumption requires that detectors be retrained whenever new technologies are released. 
As for the model, some works use classical machine learning models such as SVMs and decision trees \cite{khalid2021evaluation, borrelli2021synthetic, lataifeh2020arabic} while the majority use deep learning architectures such as DNNs \cite{wijethunga2020deepfake, zhang2021one}, CNNs \cite{camacho2021fake, liu2021identification}, and RNNs \cite{arif2021voice, shan2020cross}. 
To improve generalization to new deepfakes, some approaches try to train on a diverse set of deepfake datasets (e.g., \cite{kawa2022attack}). However, even with this strategy, ADDS systems still generalize poorly to new audio distributions recorded in new environments and to novel deepfake new technologies \cite{muller2022does}. 

In contrast to classifiers, anomaly detectors are trained on real voice data only and flag audio that has abnormal patterns within it. One approach for anomaly detection is to use the embeddings from a voice recognition model to compare the similarity between real and authentic voices \cite{pianese2022deepfake}. Other approaches use one-class machine learning models such as OC-SVMs and statistical models such as Gaussian Mixture Models (GMM) \cite{todisco2019asvspoof,zhang2021one, alegre2013one, khalid2020oc}.

What's common with the above defenses is that they are all passive defenses. This means that they analyze $a$ but they do not interact with the caller to reveal the true nature of $a$. In contrast, our proposed method is active in that it can force $f$ to try and create content it is not capable of doing. By `pressing' on the limitations of $f$, we are causing $f$ to generate audio with significantly larger artifacts, making it easier for us to detect using classifiers and anomaly detection. Our approach also ensures some longevity since the attacker cannot easily overcome the limitations our challenges pose (further discussed in section \ref{subsec:limitations}).

Another advantage of our system compared to others is that we know exactly where the anomaly should be in the media stream (due to the challenge response nature of the CAPTCHA protocol). This means that our system is more efficient since it only needs to execute its models over specific segments and not entire streams (e.g., in contrast to \cite{bartusiak2021frequency}).

The work most similar to ours is rtCAPTCHA \cite{uzun2018rtcaptcha}. In this work the authors perform liveliness detection by (1: challenge) asking the caller to read out a text CAPTCHA, (2: response) verifying that the CAPTCHA was read back correctly, and (3: robustness) verifying that the face and voice match an existing user in a database. The concept of rtCAPTCHA is that the system assumes that the attacker will not be able to generate a response with the target's face and voice in real-time. However, with the advent of RT-DFs, this rtCAPTCHA can easily be bypassed since the human attacker can read the text CAPTCHA back through $f_t$. Moreover, our D-CAPTCHA defense does not require users to register in advance, making the solution widely applicable to many users and scenarios.

\section{Deepfake CAPTCHAs}
In this section we discuss the limitations of RT-DFs and then use these limitations to define how D-CAPTCHAs work.

\subsection{RT-DF Limitations}\label{subsec:limitations}
Current RT-DF models can only generate content within the scope of the task they were trained on. For example, a model trained to reenact $t$'s face in a somewhat frontal position or generate $t$'s
voice in a calm speaking tone will not be able to generate other content. This is evident in facial reenactment models such as \cite{NIPS2019_8935} and DeepFaceLive. These models have excellent performance in creating faces with frontal poses, but they cannot generate the back of the target's head. Similarly, for audio-based RT-DFs, it is hard for the model to identify and then produce certain sounds if the training data, loss functions, and overall pipeline focuses on the perfection of normal speech. 

An \textbf{ideal RT-DF} model would be able to create content of $t$ performing an \textit{arbitrary} task, where the content is both realistic and authentic to $t$'s identity. However, RT-DF models are not ideal because they are scoped to specific tasks during training. This is because doing so enables the model to perfect the identity and realism in $x_g$ when driven by $x_s$. Therefore, even if \textit{out of domain} tasks can be anticipated, $f_t$ cannot be trained recreate them all. This is due to limitations in technology and practicality:

\subsubsection{Technology}  
    This set of limitations relates to the fact that \textit{current} technology is not yet capable of creating the \textit{ideal} RT-DF.

    \begin{description}[leftmargin=.5cm]
        \item[Inference Speed.] The rate at which audio frames can be generated depends on the efficiency of deepfake generation pipeline and the complexity of the model's architecture.
        However, in order to handle a wide variety of different tasks, a model requires significantly more parameters\footnote{As a point of reference, StarGAN \cite{StarGANv2-VC} is a state-of-the-art audio-based RT-DF models which has about 53 million parameters. In contrast, models that produce arbitrary content (such as DALL-E 2 and Imagen) use 3.5-4.6 billion parameters. Moreover, methods such as stable-diffusion requires multiple passes.} and possibly more complex feature extractors in its pipeline. For example, existing RT-DF models would need higher resolution STFTs and MFCCs to capture a wider band of frequencies. 

        \item[Feature Representation.] In order to capture certain patterns in the input $a_s$, a model must extract appropriate feature representations from the input waveform. Voice tends to use lower frequencies and has a rather consistent spectral envelope compared to other sounds such as singing and clapping. Existing pipelines use compressed features such as MFCCs or STFTs with lower sample rates (e.g., 16-24 KHz \cite{almutairi2022review}). To capture a more dynamic range of frequencies, higher resolution is needed. However, increasing input resolution generally makes it harder for a model to converge and increases model complexity.

        \item[Training.]
        To train a model, a loss function must be provided to guide the optimization process. Modern RT-DF systems use at least two loss functions: one for the realism (adversarial loss) and one for 
        preserving the identity of $t$ in $a_g$ (e.g., perceptual loss) \cite{mirsky2021creation}. If additional tasks are considered, then the model will likely need additional loss functions to cover each aspect. However, loss functions compete during optimization and therefore some aspects will suffer. Furthermore, adding loss functions can make it harder for the model to converge. Finally, it's possible that $a_s$ may contain a mix of voice and other audio (e.g., music or some other voice). To work on this audio, the model would have to convert the voice component and not the other audio, and then mix the two components back together in $x_g$. To the best of our knowledge this is an open problem.
    \end{description}

\subsubsection{Resources}
    This set of limitations relates to cases where the desired result is achievable with existing technology, however it may be prohibitively expensive or impractical to obtain it.
    
    \begin{description}[leftmargin=.5cm]
        \item[Data Collection.] To make a high quality RT-DF of $t$, a significant amount of audio samples of $t$ are required (e.g., \cite{StarGANv2-VC} requires 20-30 minutes). However, it is impractical for an attacker to obtain audio of $t$ performing specific tasks other than talking. If quality can be sacrificed, then zero-shot learning could be used. However, there is still the challenge of (1) gathering an extensive dataset of all possible tasks and (2) training a model that can generalize the samples to new identities.
        
        \item[Knowledge.] Creating a system that can handle even a subset of arbitrary tasks requires some in-depth knowledge on making generative deep learning models. This raises the difficulty bar for casual attackers, but not for advanced adversaries.

        \item[Labeling.] The process of annotating and labeling large datasets is expensive and time consuming. This becomes more apparent as the number of classes (tasks) increases. 
        
        \item[Assets.] The ideal RT-DF model would likely be a complex model to handle the arbitrary tasks. Executing such a model in real-time would require a powerful GPU. Depending on the model's complexity, the GPU may either be prohibitively expensive or simply non-existent.
    \end{description}

\subsubsection{Outlook on RT-DF Limitations}
We note that the limitations described in this section apply to existing RT-DF systems. Although these limitations are hard to overcome, there is no guarantee that future RT-DF technologies will have the same limitations. However, we expect that some of the limitations, such as data collection and training, will still apply to novel systems in the near future. 

Therefore, to gain advantage over the adversary, we suggest that defenses should exploit the limitations of RT-DFs whenever possible.

\begin{table*}[]

\caption{Examples of audio-based tasks which can be used as challenges in a D-CAPTCHA. Strong challenges are hard for the adversary on all four constraints: realism, identity, complexity and time. The measures in this list are based on existing RT-DFs methods. \textit{Playback} is where the caller must play some provided audio from his/her phone into the microphone.}\label{tab:captchas}

\resizebox{\textwidth}{!}{%
\def\arraystretch{.7}

\begin{tabular}{lccccccccc}
\multicolumn{3}{r|}{}                                                                          & \multicolumn{4}{c|}{Hardness}                                                                   & \multicolumn{1}{c|}{Weakness}          & \multicolumn{2}{c|}{Effectiveness}                                            \\
\multicolumn{1}{c}{\textbf{Task ($T$)}} & \textbf{Acronym} & \multicolumn{1}{c|}{\textbf{Usability}} & \textbf{Realism} & \textbf{Identity} & \textbf{Task} & \multicolumn{1}{c|}{\textbf{Time}} & \multicolumn{1}{c|}{\textbf{Evasions}} & \textbf{Naive   Attacker} & \multicolumn{1}{c|}{\textbf{Advanced   Attacker}} \\ \hline\hline
\textit{Clear   Throat}           & CT               & \multicolumn{1}{c|}{$\bullet$}          & $\bullet$        & $\circ$           & $\bullet$           & \multicolumn{1}{c|}{$\bullet$}     & \multicolumn{1}{c|}{}                  & $\bullet$                 & \multicolumn{1}{c|}{$\circ$}                      \\
\textit{Hold   Musical Note}      & HN               & \multicolumn{1}{c|}{$\bullet$}          & $\circ$          & $\circ$           & $\bullet$           & \multicolumn{1}{c|}{$\bullet$}     & \multicolumn{1}{c|}{}                  & $\bullet$                 & \multicolumn{1}{c|}{$\bullet$}                    \\
\textit{Hum   Tune}               & HT               & \multicolumn{1}{c|}{$\bullet$}          & $\bullet$        & $\bullet$         & $\bullet$           & \multicolumn{1}{c|}{$\bullet$}     & \multicolumn{1}{c|}{}                  & $\bullet$                 & \multicolumn{1}{c|}{$\bullet$}                    \\
\textit{Laugh}                    & L                & \multicolumn{1}{c|}{$\circ$}            & $\bullet$        & $\bullet$         & $\bullet$           & \multicolumn{1}{c|}{$\bullet$}     & \multicolumn{1}{c|}{}                  & $\bullet$                 & \multicolumn{1}{c|}{$\bullet$}                    \\
\textit{Mimic   Speaking Style}   & MS               & \multicolumn{1}{c|}{$\circ$}            & $\bullet$        & $\bullet$         & $\circ$             & \multicolumn{1}{c|}{$\bullet$}     & \multicolumn{1}{c|}{}                  & $\circ$                   & \multicolumn{1}{c|}{$\circ$}                      \\
\textit{Repeat   Accent}          & R                & \multicolumn{1}{c|}{$\circ$}            & $\bullet$        & $\bullet$         & $\circ$             & \multicolumn{1}{c|}{$\bullet$}     & \multicolumn{1}{c|}{}                  & $\circ$                   & \multicolumn{1}{c|}{$\circ$}                      \\
\textit{Sing}                     & S                & \multicolumn{1}{c|}{$\bullet$}          & $\bullet$        & $\bullet$         & $\bullet$           & \multicolumn{1}{c|}{$\bullet$}     & \multicolumn{1}{c|}{}                  & $\bullet$                 & \multicolumn{1}{c|}{$\bullet$}                    \\
\textit{Speak   with Emotion}     & SE               & \multicolumn{1}{c|}{$\bullet$}          & $\bullet$        & $\bullet$         & $\circ$             & \multicolumn{1}{c|}{$\bullet$}     & \multicolumn{1}{c|}{}                  & $\bullet$                 & \multicolumn{1}{c|}{$\bullet$}                    \\
\textit{Yawn}                     & Y                & \multicolumn{1}{c|}{$\circ$}            & $\bullet$        & $\circ$           & $\bullet$           & \multicolumn{1}{c|}{$\bullet$}     & \multicolumn{1}{c|}{}                  & $\bullet$                 & \multicolumn{1}{c|}{$\bullet$}                    \\
\textit{Blow   Noises}            & BN               & \multicolumn{1}{c|}{$\bullet$}          & $\bullet$        & $-$               & $\bullet$           & \multicolumn{1}{c|}{$\bullet$}     & \multicolumn{1}{c|}{bypass}            & $\bullet$                 & \multicolumn{1}{c|}{$-$}                          \\
\textit{Blow   on Mic}            & BM               & \multicolumn{1}{c|}{$\circ$}            & $\bullet$        & $-$               & $\bullet$           & \multicolumn{1}{c|}{$\bullet$}     & \multicolumn{1}{c|}{bypass}            & $\bullet$                 & \multicolumn{1}{c|}{$-$}                          \\
\textit{Clap}                     & Cl               & \multicolumn{1}{c|}{$\bullet$}          & $\circ$          & $-$               & $\bullet$           & \multicolumn{1}{c|}{$\bullet$}     & \multicolumn{1}{c|}{bypass}            & $\bullet$                 & \multicolumn{1}{c|}{$-$}                          \\
\textit{Click   Tongue}           & Clk              & \multicolumn{1}{c|}{$\bullet$}          & $\bullet$        & $-$               & $\bullet$           & \multicolumn{1}{c|}{$\bullet$}     & \multicolumn{1}{c|}{bypass}            & $\bullet$                 & \multicolumn{1}{c|}{$-$}                          \\
\textit{Cough}                 & Co               & \multicolumn{1}{c|}{$\bullet$}          & $\bullet$        & $-$               & $\bullet$           & \multicolumn{1}{c|}{$\bullet$}     & \multicolumn{1}{c|}{bypass}            & $\bullet$                 & \multicolumn{1}{c|}{$-$}                          \\
\textit{Horse   Lips}             & HL               & \multicolumn{1}{c|}{$\circ$}            & $\bullet$        & $-$               & $\bullet$           & \multicolumn{1}{c|}{$\bullet$}     & \multicolumn{1}{c|}{bypass}            & $\bullet$                 & \multicolumn{1}{c|}{$-$}                          \\
\textit{Knock}                    & K                & \multicolumn{1}{c|}{$\circ$}            & $\circ$          & $-$               & $\bullet$           & \multicolumn{1}{c|}{$\bullet$}     & \multicolumn{1}{c|}{bypass}            & $\bullet$                 & \multicolumn{1}{c|}{$-$}                          \\
\textit{Playback   Audio}         & PA                & \multicolumn{1}{c|}{$-$}                & $\bullet$        & $-$               & $\bullet$           & \multicolumn{1}{c|}{$\bullet$}     & \multicolumn{1}{c|}{bypass}            & $\bullet$                 & \multicolumn{1}{c|}{$-$}                          \\
\textit{Raspberry}                & R                & \multicolumn{1}{c|}{$\bullet$}          & $\bullet$        & $-$               & $\bullet$           & \multicolumn{1}{c|}{$\bullet$}     & \multicolumn{1}{c|}{bypass}            & $\bullet$                 & \multicolumn{1}{c|}{$-$}                          \\
\textit{Sound   Effect}           & SFX              & \multicolumn{1}{c|}{$\bullet$}          & $\bullet$        & $-$               & $\bullet$           & \multicolumn{1}{c|}{$\bullet$}     & \multicolumn{1}{c|}{bypass}            & $\bullet$                 & \multicolumn{1}{c|}{$-$}                          \\
\textit{Touch   Mic}              & TM               & \multicolumn{1}{c|}{$\circ$}            & $\bullet$        & $-$               & $\bullet$           & \multicolumn{1}{c|}{$\bullet$}     & \multicolumn{1}{c|}{bypass}            & $\bullet$                 & \multicolumn{1}{c|}{$-$}                          \\
\textit{Type}                     & T                & \multicolumn{1}{c|}{$\circ$}            & $\bullet$        & $-$               & $\bullet$           & \multicolumn{1}{c|}{$\bullet$}     & \multicolumn{1}{c|}{bypass}            & $\bullet$                 & \multicolumn{1}{c|}{$-$}                          \\
\textit{Whistle}                  & W                & \multicolumn{1}{c|}{$-$}                & $\bullet$        & $-$               & $\bullet$           & \multicolumn{1}{c|}{$\bullet$}     & \multicolumn{1}{c|}{bypass}            & $\bullet$                 & \multicolumn{1}{c|}{$-$}                          \\
\textit{Talk   \& Clap}           & T\&C             & \multicolumn{1}{c|}{$\circ$}            & $\bullet$        & $\bullet$         & $\bullet$           & \multicolumn{1}{c|}{$\bullet$}     & \multicolumn{1}{c|}{mix}            & $\bullet$                 & \multicolumn{1}{c|}{$-$}                          \\
\textit{Talk   \& Knock}          & T\&K             & \multicolumn{1}{c|}{$\circ$}            & $\bullet$        & $\bullet$         & $\bullet$           & \multicolumn{1}{c|}{$\bullet$}     & \multicolumn{1}{c|}{mix}            & $\bullet$                 & \multicolumn{1}{c|}{$-$}                          \\
\textit{Talk   \& Playback}       & P             & \multicolumn{1}{c|}{$-$}                & $\bullet$        & $\bullet$         & $\bullet$           & \multicolumn{1}{c|}{$\bullet$}     & \multicolumn{1}{c|}{mix}            & $\bullet$                 & \multicolumn{1}{c|}{$-$}                          \\
\textit{Talk   with Tones}        & TT               & \multicolumn{1}{c|}{$\bullet$}          & $\bullet$        & $\bullet$         & $\bullet$           & \multicolumn{1}{c|}{$\bullet$}     & \multicolumn{1}{c|}{mix}            & $\bullet$                 & \multicolumn{1}{c|}{$\bullet$}                    \\
\textit{Vary   Speed}             & VS               & \multicolumn{1}{c|}{$\bullet$}          & $\bullet$        & $\bullet$         & $\circ$             & \multicolumn{1}{c|}{$\bullet$}     & \multicolumn{1}{c|}{mix}            & $\bullet$                 & \multicolumn{1}{c|}{$\bullet$}                    \\
\textit{Vary   Volume}            & V                & \multicolumn{1}{c|}{$\bullet$}          & $\bullet$        & $\bullet$         & $\circ$             & \multicolumn{1}{c|}{$\bullet$}     & \multicolumn{1}{c|}{mix}            & $\bullet$                 & \multicolumn{1}{c|}{$\bullet$}                    \\ \hline\hline
\multicolumn{10}{l}{$\bullet$: high, $\circ$: medium, $-$: low}                                                                                                                                                                                                                                                          
\end{tabular}%
}
\vspace{-1em}
\end{table*}

\subsection{D-CAPTCHA}

According to \cite{ahn2003captcha}, a CAPTCHA is ``\textit{a cryptographic protocol whose underlying hardness assumption is based on an AI problem.}'' The protocol follows the form of a challenge-response procedure between server $A$ (the server/victim) and client $B$ (the client/caller), where (1) $A$ sends challenge $c$ to $B$, (2) $B$ sends response $r_c$ on $c$ back to $A$, and (3) $A$ verifies whether $r_c$ resolves challenge $c$:
\begin{enumerate}
    \item $A\rightarrow B: c$
    \item $B\rightarrow A: r_c$
    \item $A: V(r_c) \in \{pass, fail\}$
\end{enumerate}
For example, the popular reCAPTCHA prevents bots from performing automated activities on the web by challenging the client to perform a human skill which is hard for software but easy for humans (e.g., decoding distorted letters).
In contrast, a D-CAPTCHA challenges a client by requiring the client to \textbf{create} content with the following constraints:
\begin{enumerate}
    \item \textbf{Realism}: The content must be realistic to a human or a machine learning model
    \item \textbf{Identity}: The content must reflect the identity $t$
    \item \textbf{Task}: The content must have $t$ performing an arbitrary task which is hard to generate
    \item \textbf{Time}: The content must be generated in real-time
\end{enumerate}
Creating a response to this challenge where $V(r_c)=pass$ is hard for existing RT-DF technologies but easy for humans. In our system the `hardness' of the CAPTCHA directly relates to the limitations of existing RT-DF technology (section \ref{subsec:limitations}). Moreover, just like modern CAPTCHA systems, a D-CAPTCHA system can be easily extended to new limitations of RT-DFs over time. This gives our system flexibility to defend against future threats.

\subsubsection{Creating a Challenge}
A challenge demonstrates whether a caller can or cannot create content with realism, identity, task and time constraints. \textbf{Realism} constraints are necessary to ensure there are no latent or semantic anomalies in the response. \textbf{Identity} constraints are needed to ensure that the attacker isn't just recording him/herself during the challenge. \textbf{Task} constraints are required to ensure that the deepfake model tries to operate outside the bounds of its abilities. Finally, \textbf{Time} constraints are involved to guarantee that the caller is using an RT-DF model since (1) we don't want the caller to switch to an offline model and (2) real-time models are more limited since they can only process frames and not entire audio clips.

The core component of a challenge in our system is the task which the caller must perform. Let $T$ denote a specific task, such that $T=hum$ might be ``hum a specific song.'' We define the set of all possible challenges for task $T$ as $C_T$. For example,  $C_{hum}$ would be all possible requests for different songs to be hummed. To select a challenge, (1) random seeds $z_0, z_1$ are generated, (2) $z_0$ is used to select a random task $T$ and (3) $z_1$ is used to select a random challenge $c \in C_T$.

In Table \ref{tab:captchas} we present some example tasks which can be used in D-CAPTCHA challenges. In the table, we assume that the RT-DF under test has been trained to have the best performance on one task; regular talking.
Using observations over five state-of-the-art RT-DF models we assess the hardness, weakness, and effectiveness of each task as a challenge  (see \ref{subsubsec:RTDF} for details on these five models). Under hardness, we express the difficulty of a modern RT-DF in successfully creating a deepfake of $t$ given the respective constraints. For weakness, we state how an adversary can evade detection if the respective task is chosen. For instance, \textit{bypass} is where the RT-DF is turned off and the attacker speaks directly to our system. The other case is \textit{mix} is where the attacker can mix other audio sources into $a_g$. For example, to evade `talk \& clap' the attacker creates $a'_g = a_g + a_{clap}$ where $a_{clap}$ is taken from another microphone so as not to disrupt the RT-DF (i.e., execute $f_t(a_s + a_m)$). Finally, in the table under effectiveness we consider how effective the challenge is given two levels of attackers: naive and advanced. A naive attacker is one which (1) will use existing datasets and only a limited number of samples of $t$ to train $f_t$ and (2) forwards all audio through $f_t$ (e.g., if a library is used as-is from GitHub). An advanced attacker is one which will collect a practical number of samples on $t$ (e.g., 20 minutes) and is able to mix other audio sources into $a_g$. 

Overall, a strong challenge is a random $c$ drawn from a random $T$ which is hard for the adversary to perform given all four constraints.

\begin{figure*}[t]
    \centering
    \includegraphics[width=\textwidth]{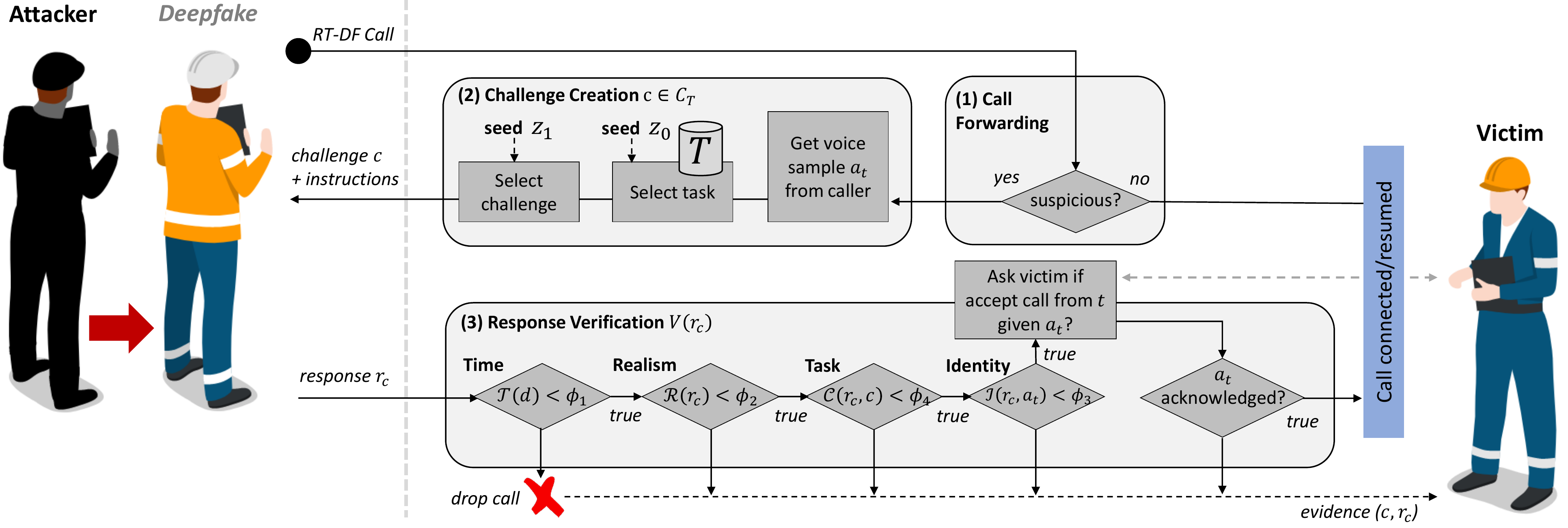}
    \caption{An overview of the proposed D-CAPTCHA system: (1) Calls are forwarded to the system using a blacklist, whitelist, policy or the victim's intuition, (2) a random D-CAPTCHA $c$ with accompanying instructions is generated and send to the caller as a challenge, (3) the response $r_c$ is verified against the four constraints (time, realism, identity, task) and if all four pass then the call is connected/resumed. Otherwise, the call is dropped and evidence is provided to the victim.}
    \label{fig:framework}
\end{figure*}

\subsubsection{Verifying a Challenge}\label{subsubsec:verify}
To determine whether $V(r_c)=pass$ or $fail$, we must verify whether $r_c$ adheres to the \textbf{realism}, \textbf{identity}, \textbf{Task}, and \textbf{time} constraints. All four constraints can be verified by a human (a moderator or the victim him/herself). For example, if $c=$``say 'I'm hungry' with anger'' but (1) the audio sounds strange/distorted, (2) the voice does not sound like $t$, (3) the task is not completed, or (4) it takes too long for the caller to respond, then this would raise suspicion. However, many users may not trust themselves enough or they may give in to social pretexts and ignore the signs --to avoid rejecting a peer. Therefore, we propose an automated way to verify each constraint without prior knowledge of $t$.

\noindent To verify $r_c$, we validate each constraint separately:
\begin{description}[leftmargin=.5cm]
    \item[Realism Verification ($\mathcal{R}$).] If an RT-DF attempts to perform $c$ then $r_c$ will likely contain distortions and artifacts. This is because (1) the RT-DF is operating outside of its capabilities or (2) because the caller simply is using a poor-quality RT-DF. These distortions will make it easier for existing anomaly detectors and existing deepfake classifiers to identify the RT-DF. The output of $\mathcal{R}$ is a score on the range $[0,\infty)$ or $[0,1]$ indicating how unrealistic the content of $r_c$ is.
    
    \item[Identity Verification ($\mathcal{I}$).] To determine if $r_c$ has the identity $t$, we can do as follows: (1) collect a short audio sample $a_t$ of the caller prior to the challenge and have the victim acknowledge the identity, and (2) use zero-shot voice recognition model to verify that the identity in $a_t$ and $r_c$ are the same. The reason we have the victim acknowledged $t$ in $a_t$ is to prevent the attacker from switching the identity after the challenge. Alternatively, interaction with the victim can be avoided if continuous voice verification is used on the caller. However, doing so would be expensive. The output of $\mathcal{I}$ is a similarity score between $a_t$ and $r_c$.
    
    \item[Task Verification ($\mathcal{C}$).] There are two cases where $r_c$ would not contain the requested task: (1) the model failed to generate the content and (2) the attacker is trying to evade generating artifacts by performing another task or nothing at all. To ensure that $r_c$ contains the task, we can use a machine learning classifier. The output of $\mathcal{C}$ is the probability that $r_c$ \textit{does not} contain the task.
 
    \item[Time Verification ($\mathcal{T}$).] The time constraint can be verified by ensuring that the first frame of $r_c$ is received within roughly $1$ second after of the challenge's start time (i.e., after the instructions for $c$ are given). The output of $\mathcal{T}$ is the measured time delay denoted $d$.
\end{description}

Altogether, we validate $r_c$ if none of the four algorithms ($\mathcal{T},\mathcal{R},\mathcal{I},\mathcal{C}$) exceed their respective thresholds ($\phi_1,\phi_2,\phi_3,\phi_4$) where each threshold has been tuned accordingly. We invalidate $r_c$ if any model exceeds its respective threshold. The false reject rate can be tuned by weighing the contribution of each constraint, however doing so will compromise the security of the system.

In summary, validation is performed as follows:
\begin{equation}
    V(r_c) = 
    \begin{cases}
      pass, \hspace{1em} \mathcal{T}(d)< \phi_1, \mathcal{R}(r_c)< \phi_2, \\
      \hspace{3em} \mathcal{I}(r_c,a_t)< \phi_3, \mathcal{C}(r_c,c) < \phi_4    \\
      fail, \hspace{1em} \text{else}
    \end{cases}
\end{equation}
We note that a combination of validation methods for each constraint can be used to increase performance, security and usability. For example, some verifications can be done with humans, some with algorithms and some with both.

\section{Detection Framework}

In this section we present the D-CAPTCHA framework which can be used to protect users (victims) from fake callers. A summary of the D-CAPTCHA framework can be found in Fig. \ref{fig:framework}.

\subsection{1: Call Forwarding}

The very first step is to decide which calls should be forwarded to the system. In high risk settings, a D-CAPTCHA may be used to verify every caller. However, this is not practical in most settings. Instead, calls can be forwarded to the system using blacklists (e.g., known offenders) or policies. An example policy is to forward all callers who are not in the victim's address book, or to screen all calls during working hours. 

Alternatively, call screening can be activated by the user. For example, if a call arrives from an unknown number, the user can choose to forward it to the D-CAPTCHA system if the call is unexpected. Another option is to let users forward ongoing calls if (1) the caller's audio sounds strange, (2) the conversation is suspicious, or (3) a sensitive discussion needs to be made. For example, consider the scenario where a user receives a call from a friend under an odd pretext such as ``I'm stuck in Brazil and need money to get out.'' Here, the user can increase his/her confidence in the caller's authenticity after forwarding the call through the D-CAPTCHA system. 

\subsection{2: Challenge Creation}

A random challenge $c$ is generated using the approach described in section \ref{subsubsec:verify}. In addition to $c$, instructions for the caller are generated. Instructions include a list of actions to perform and a start indicator. For example, an instruction might be ``at the tone, knock three times while introducing yourself.'' The instruction is then converted into an audio message using TTS.

At the start of the challenge, the caller is asked to state his/her name. This recording is saved as $a_t$ and shared with the victim for acknowledgment and with $\mathcal{I}$ for identity verification.\footnote{Recall, this is done to prevent attackers from simply turning off the RT-DF during the challenge and using their actual voice.} Next, the audio instructions are played to the caller. After playing the instructions, a tone is sounded. The time between the tone and the first audible sounds from the caller is measured and included as part of $r_c$ for $\mathcal{T}$. After a set number of seconds, the caller's recording is saved as $r_c$ and passed along for verification.


\subsection{3: Response Verification}

The recorded response $r_c$ and its timing data are sent to $\mathcal{T}, \mathcal{R}, \mathcal{C},$ and $\mathcal{I}$ for constraint verification. If all the algorithms yield scores below their respective thresholds, then $a_t$ is played to the user. If the user accepts the call with $t$ then the D-CAPTCHA is $valid$ and the call is connected / resumed. 

If any of the algorithms produce a score above their threshold, then the call is dropped, and evidence is provided to the user. Evidence consists of an explanation of why the call was not trusted (e.g., information on which constraint(s) failed and to what degree) and playback recordings of $a_t$, $c$, and $r_c$ accordingly. Although the order which the models are executed does not matter, we can avoid executing redundant models if one model detects the deepfake. Therefore, we suggest the order $\mathcal{T}\rightarrow\mathcal{R}\rightarrow\mathcal{C}\rightarrow\mathcal{I}$ to potentially save execution time when detecting a deepfake. We also note that if higher security is required, then multiple D-CAPTCHAs can be sent out and subsequently verified to reduce the false negative rate. 









\subsubsection{Deployment}
In general, the framework can be deployed as an app on the victim's phone or as a service in the cloud. For example, onsite technicians, bankers, and the elderly can have the system screen calls directly on their phones. Call centers and online meeting rooms can use cloud resources to screen callers in waiting rooms (e.g., before connecting to a confidential Zoom meeting \cite{European87:online,Binancee98:online}).

\subsection{Limitations}
The main limitations of this system are its applicability and usability. In terms of deployment, the system must be able to interact with the deepfake so it can only protect against RT-DFs. Moreover, since it is an active defense, the CAPTCHA protocol runs the risk of becoming a hindrance to users if not tuned correctly. Regardless, it's a great solution for screening callers entering high security conversations and meetings in an age where calls cannot be trusted. Finally, the system uses deep learning models in $\mathcal{R}$, $\mathcal{I}$, and $\mathcal{C}$. Just like other deep learning-based defenses, an attacker can potentially evade these models using adversarial examples \cite{carlini2020evading}. However, when trying to evade our system, the attacker must overcome a number of challenges: (1) most calls are made over noisy and compressed channels reducing the impact of the perturbations, (2) performing this attack would require real-time generation of adversarial examples, and (3) $\mathcal{R}$, $\mathcal{I}$, and $\mathcal{C}$ would most likely be a black box to the attacker, although not impervious, it cannot be easily queried.

\section{Threat Analysis} 
In this section, we assess the threat posed by RT-DFs by evaluating the quality of five state-of-the-art RT-DF models in the perspective of 41 volunteers. 

\subsection{Experiment Setup}
\subsubsection{RT-DF Models}\label{subsubsec:RTDF} 
We surveyed 25 voice cloning papers published over the last three years which can process audio in real-time as a sequence of frames. Of the 25 papers we selected the four recent works which published their source code: \texttt{AdaIN-VC}  \cite{AdaIN}, \texttt{MediumVC} \cite{MediumVC}, \texttt{FragmentVC} \cite{FragmentVC} and \texttt{StarGANv2-VC}  \cite{StarGANv2-VC}. We also selected \texttt{ASSEM-VC} \cite{Assem-VC} which is a non-casual model as an additional comparison. All works are from 2021 except \texttt{AdaIN-VC} which is from 2019.

\begin{description}[leftmargin=.5cm]
    \item[any-to-many.] \texttt{StarGANv2-VC} is many-to-many model which also works as an any-to-many model. The audio $a_g$ is created by passing the spectrogram of $a_s$ through an encoder-decoder network. To disentangle content from identity, the decoder also receives an encoding of $a_s$ taken from a pretrained network which extracts the fundamental frequencies. Finally, the decoder receives reference information on $t$ via a style encoder using sample $a_t$. \texttt{ASSEM-VC} works in a similar manner except $a_s$ and a TTS transcript of $a_s$ are used to generate a speaker independent representation before being passed to the decoder, and the decoder receives reference information on $t$ from an identify encoder.

    \item[any-to-any.] In \texttt{AdaIN-VC}, $a_g$ is created by disentangling identity from content. The model (1) passes a sample $a_t$ through an identity encoder, (2) passes a source frame $a_s^{(i)}$ through a content encoder with instance-normalization, and then (3) passes both outputs through a final decoder. In \texttt{MediumVC}, $a_s$ first normalizes the voice by converting it to a common identity with an any-to-one VC model. The result is then encoded and passed to a decoder along with an identity encoding (similar to \texttt{AdaIN-VC}). \texttt{FragmentVC}, extracts the content of $a_s$ using a Wav2Vec 2.0 model \cite{wav2vec} and extracts fragments of $a_t$ using an encoder. A decoder then uses attention layers to fuse the identity fragments into the content to produce $a_g$.
\end{description}
All audio clips in this experiment were generated using the pre-trained models provided by the original authors. To simulate a realistic setting, the clips were passed through a phone filter (a band pass filter on the 0.3-3KHz voice range).

\subsubsection{Experiments} 
To help quantify the threat of RT-DFs, we performed two experiments on a group of 41 volunteers:
\begin{description}[leftmargin=.5cm]
    \item[EXP1a - Quality.] The goal of the first experiment was to see how easy it is to identify an RT-DF in the best-case scenario (when the victim is expecting a deepfake).
    \item[EXP1b - Identity.] The goal of this experiment was to understand how well RT-DF models are able to clone identities.
\end{description}

In \textbf{EXP1a}, volunteers were asked to rate audio clips on a scale of 1-5 (1: fake, 5: real). There were 90 audio clips presented in random order: 30 real and 60 fake (12 from each of the five models). The clips were about 4-7 seconds long each. 

In \textbf{EXP1b}, we selected the top 2 models that performed the best in EXP1. For each model, we repeated the following trial 8 times: We first let the volunteer listen to two real samples of the target identity as a baseline. Then we played two real and two fake samples in random order and asked the volunteer to rate how similar their speakers sound compared to the speaker in the baseline. 

If a model has a positive mean opinion score (MOS) in both EXP1 and EXP2 then it is a considerable threat. This is because it can (1) synthesize high quality speech (2) that sounds like the target (3) all in real-time.

\subsection{Experiment Results}

\begin{figure}
    \centering
    \includegraphics[width=.95\columnwidth]{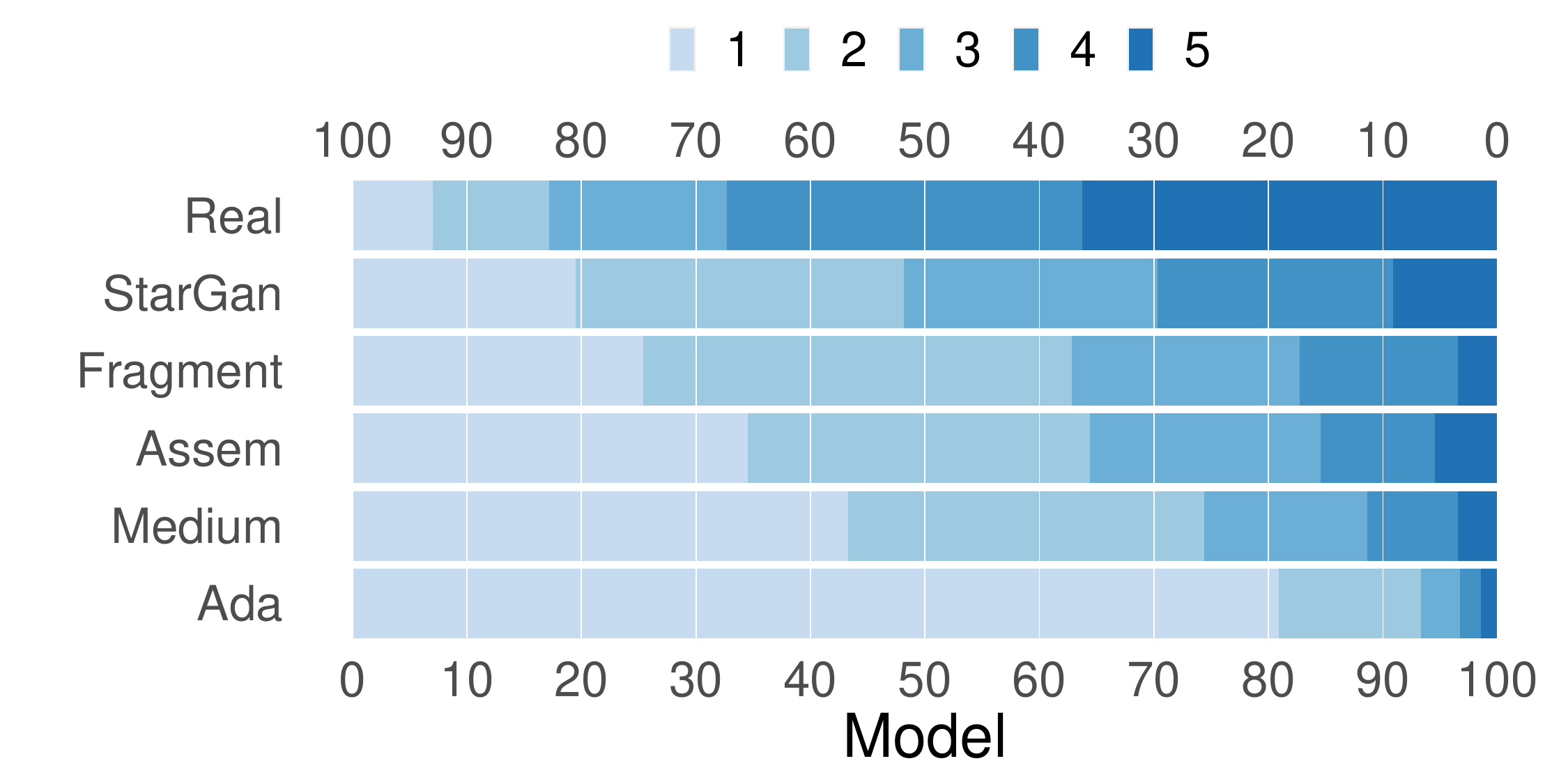}
    \caption{RT-DF Quality - The distribution of ratings which the volunteers gave to each of the RT-DF models and real voice recordings (1: fake, 5: real). }
    \label{fig:exp1.1}

    \centering
    \includegraphics[width=\columnwidth]{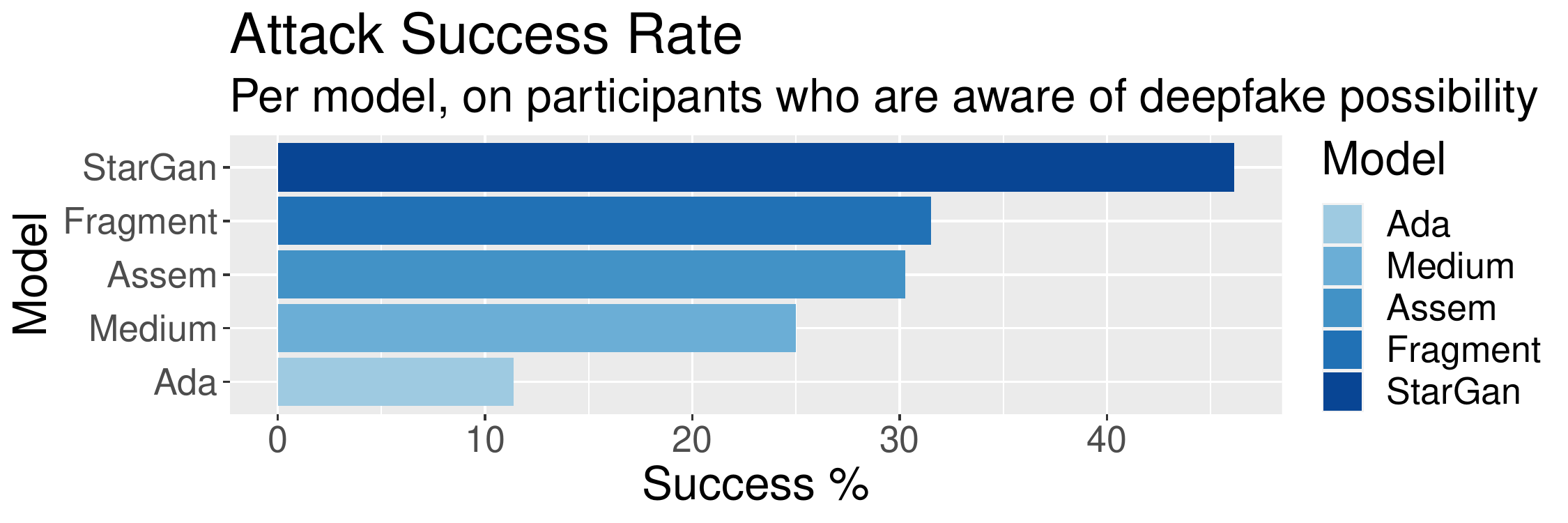}
    \caption{RT-DF Quality - The percent of volunteers fooled by each RT-DF model, even though they were expecting a deepfake.}
    \label{fig:exp1.2}

    \centering
    \includegraphics[width=\columnwidth]{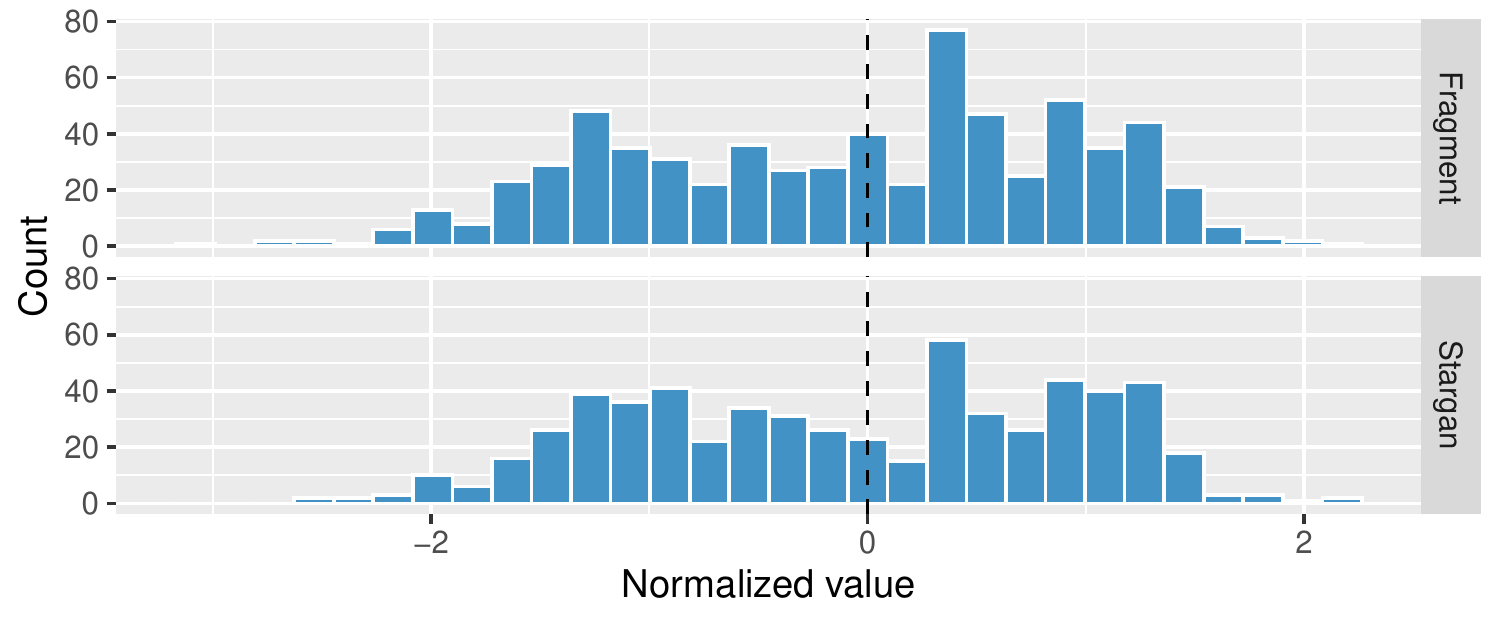}
    \caption{RT-DF Identity - A histogram of the normalized MOS scores for how similar RT-DF audio sounds like the target identity $t$. Positive scores are cases where volunteers thought a fake audio sounded more like $t$ than an authentic recording of $t$. }
    \label{fig:exp2.1}
\end{figure}

\textbf{EXP1a.} To analyze the quality (realism) of the models, we compared the MOS scores of the deepfake audio to the MOS of the real audio (both scored blindly). In Fig. \ref{fig:exp1.1} we plot the distribution of each model's MOS compared to real audio. Roughly 20-50\% of the volunteers gave the RT-DF audio positive score with \texttt{StarGANv2-VC} having the highest quality. 

However, opinion scores are subjective. Therefore, we need to normalize the MOS to count how many times volunteers were fooled by an RT-DF. In principle, the range of scores a volunteer $k$ has given to real audio captures that volunteer's `trust' range. Let $\mu_{real}^k$ and $\sigma_{real}^k$ be the mean and standard deviation on $k$'s scores for real clips. We estimate that a volunteer would likely be fooled by a clip if he or she scores a clip with a value greater than $\mu_{real}^k - \sigma_{real}^k$. 

Using this measure, in Fig. \ref{fig:exp1.2} we present the attack success rate for each of the RT-DF models. We found that \texttt{StarGANv2-VC} has the highest success rate of 46\% percent rate. This means that although current RT-DF models are not perfect, they can indeed fool people. We note that these results cannot be interpreted as the likelihood of a true RT-DF attack succeeding. This is because our volunteers were expecting to hear deepfakes and were therefore carefully listening for artifacts. A true victim would likely overlook some artifacts especially when put under pressure by the attacker. 

\textbf{EXP1b.} To analyze the ability of the models to copy identities, we normalized volunteer $k$'s scores on \textit{fake} audio by computing $\frac{score - \mu_{real}^k}{\sigma_{real}^k}$. Fig. \ref{fig:exp2.1} plots the distribution of the normalized scores on fake audio. We can see that the volunteers were mostly indecisive, rating some fake clips as more authentic and some as less. For the majority of cases ($score > -1)$ volunteers felt that the identity was captured well by the top two models.

In summary, there is a chronological trend given that the worst performing model \texttt{AdaIN-VC} is from 2019 and the best \texttt{StarGANv2-VC} is from 2021. This may indicate that the quality of RT-DF is rapidly improving. This raises concern, especially since the volunteers were expecting the attack yet could not accurately tell which clips were real or fake. Another insight we have is that the presence of artifacts can help victims identify RT-DFs. However, as quality improves, we expect that only way to induce significant artifacts will be by challenging the model.

\section{D-CAPTCHA Evaluation}\label{sec:eval}
In this section, we evaluate the benefit of using a D-CAPTCHA as opposed to using passive defenses alone. 

\subsection{Experiment Setup}

\subsubsection{Datasets}
To evaluate our system, we recorded 20 English speaking volunteers to create both speech and challenge-response datasets, summarized in Table \ref{tab:datasets}:
\begin{description}[leftmargin=.5cm]
    \item[($\mathcal{D}_{real}$)] 2498 samples of real speech (100-250 random sentences spoken by each of the 20 volunteers).
    
    \item[($\mathcal{D}_{fake}$)] 1821 samples of RT-DF voice conversion. To create this dataset we used \texttt{StarGANv2-VC} which was the top performing model from \textbf{EXP1a}. The model was trained to impersonate 6 of the 20 volunteers from $\mathcal{D}_{real}$, and an additional 14 random voice actors from the VCTK dataset. The additional 14 were added to help the model generalize better, and only the 6 volunteers' voices were used to make RT-DFs.
    
    \item[($\mathcal{D}_{real,r}$)] 3317 samples of real responses (attempts at challenges). A sample of nine tasks were evaluated in total. The following tasks were performed ~30 times per volunteer: sing (S), hum tune (HT), coughing (Co), vary volume (V), and talk \& playback (P), and the following tasks were performed ~5 times per volunteer: repeat accent (R), clap (Cl), speak with emotion (SE), and vary speed (VS).
    
    \item[($\mathcal{D}_{fake,r}$)] 16,123 deepfake samples of RT-DF voice conversion applied to the responses $\mathcal{D}_{real,r}$ using \texttt{StarGANv2-VC}. We did not convert samples from the same identity (i.e., where $s=t$)
\end{description}
It took each volunteer over an hour to record their data. The volunteers were compensated for their time. For all train-test splits used in our evaluations, we made sure not to use the same identities in both the train and test sets. 

In addition, we also used public deepfake datasets to train the realism models $\mathcal{R}$. These datasets were the ASVspoof-DF dataset \cite{yamagishi2021asvspoof} with 22,617 real and 15,000 fake samples, and the RITW dataset \cite{muller2022does} with 19,963 real and 11,816 fake samples. 

\subsubsection{Models}
Our system, when fully automated, consists of 3 models: $\mathcal{R}$, $\mathcal{C}$ and $\mathcal{I}$. The algorithm $\mathcal{T}$ does not use a machine learning model to verify the time constraint.

For the realism model $\mathcal{R}$, we evaluated five different deepfake detection models: \textbf{SpecRNet} \cite{kawa2022specrnet} which is a novel neural network architecture, inspired by RawNet2 \cite{rawnet2}, which get results comparable to state–of–the-art models despite a significant decrease in computational requirements. \textbf{One-Class} \cite{zhang2021one} is a method adapted from \cite{monteiro2020generalized} based on a deep residual network ResNet-18 \cite{he2016deep}. They improve and generalize the network performance using One-Class Softmax activations. \textbf{GMM-ASVspoof} \cite{yamagishi2021asvspoof} is a Gaussian mixture model (GMM) which operates on LFCCs features. This model was a baseline for the in ASVspoof 2021 competition. \textbf{PC-DARTS} \cite{ge2021raw} is a convolutional neural network (CNN) that tries to automatically learn the network's architecture. This work also showed good results in generalizing to unseen attacks. Finally, we used \textbf{Local Outlier Factor} (LOF) which is a density-based anomaly detection model.

We took the union of ASVspoof-DF and RITW and selected 80\% at random for training the models and 10\% for validation (early stopping). The models were tested on the baseline scenario ($\mathcal{D}_{real}$ and $\mathcal{D}_{fake}$) and our proposed D-CAPTCHA scenario ($\mathcal{D}_{real,r}$ and $\mathcal{D}_{fake,r}$).

For the task model $\mathcal{C}$, we trained a GMM classifier on the MFCC features using the baseline model from \cite{yamagishi2021asvspoof}. One model was trained per task: to classify between real responses from that task and all other tasks as well as speech. A 70-30 train-test split was used.

For the identity model $\mathcal{I}$, we used a pre-trained voice recognition model from the SpeechBrain toolkit \cite{speechbrain}. The model uses the ECAPA-TDNN architecture to classify a speaker. Since we do not want $\mathcal{I}$ to have prior knowledge of $t$, we converted the model into an anomaly detector. Recall that we obtain a voice sample $a_t$ from the caller prior to the challenge. This sample is used as a reference to ensure that the RT-DF is not turned off during the challenge. To detect whether the identity of the caller has changed during the challenge, we compute
\begin{equation}
   \mathcal{I}(a_t, r_c)  = ||f^*(a_t)-f^*(r_s)||^2
\end{equation}
where $f^*$ is the speaker encoding, taken from an inner layer of the speech recognition model. Smaller scores indicate similarity between the voice before the challenge and during the challenge. This technique of comparing speaker encodings has been done in the past (e.g., \cite{pianese2022deepfake, mirsky2021creation}). To evaluate $\mathcal{I}$, we create negative pairings as samples from the same identity $(a_i, r_{c,i})$ and positive pairings as samples from different identities $(a_i, r_{c,j})$, where\\ $a_i,a_j\in\mathcal{D}_{real}$, \hspace{0.5em} $r_{c,i},r_{c,j}\in\mathcal{D}_{real,r}$ \hspace{0.5em} and $i\neq j$.

\begin{table}[]
\caption{The number of samples in each of our datasets}
\label{tab:datasets}
\resizebox{0.9\columnwidth}{!}{%
\def\arraystretch{.9}

\begin{tabular}{rcc}
\multicolumn{1}{r|}{\textit{}}                        & \textbf{Real: $\mathcal{D}_{real}$}   & \multicolumn{1}{c|}{\textbf{Fake: $\mathcal{D}_{fake}$}}   \\ \cline{2-3} 
\multicolumn{1}{r|}{\textit{Speech}}                  & 2498                                 & \multicolumn{1}{c|}{1821}                                 \\ \cline{2-3} 
\textit{}                                             &                                        &                                                             \\
\multicolumn{1}{r|}{\textit{}}                        & \textbf{Real: $\mathcal{D}_{real,r}$} & \multicolumn{1}{c|}{\textbf{Fake: $\mathcal{D}_{fake,r}$}} \\ \cline{2-3} 
\multicolumn{1}{r|}{\textit{Repeat Accent (R)}}       & 98                                     & \multicolumn{1}{c|}{570}                                    \\
\multicolumn{1}{r|}{\textit{Clap (Cl)}}               & 99                                     & \multicolumn{1}{c|}{551}                                    \\
\multicolumn{1}{r|}{\textit{Cough (Co)}}              & 537                                    & \multicolumn{1}{c|}{3,401}                                  \\
\multicolumn{1}{r|}{\textit{Speak with Emotion (SE)}} & 98                                     & \multicolumn{1}{c|}{532}                                    \\
\multicolumn{1}{r|}{\textit{Hum Tune (HT)}}           & 593                                    & \multicolumn{1}{c|}{3,325}                                  \\
\multicolumn{1}{r|}{\textit{Playback Audio (P)}}      & 601                                    & \multicolumn{1}{c|}{3,420}                                  \\
\multicolumn{1}{r|}{\textit{Sing (S)}}                & 595                                    & \multicolumn{1}{c|}{334}                                    \\
\multicolumn{1}{r|}{\textit{Vary Speed (VS)}}         & 98                                     & \multicolumn{1}{c|}{570}                                    \\
\multicolumn{1}{r|}{\textit{Vary Volume (V)}}         & 598                                    & \multicolumn{1}{c|}{3,420}                                  \\ \cline{2-3}  
\textit{}                                             &                                        &                                                             \\

\multicolumn{1}{r|}{\textit{}}                        & \textbf{Real}   & \multicolumn{1}{c|}{\textbf{Fake}}   \\ \cline{2-3} 
\multicolumn{1}{r|}{\textit{ASVspoof-DF}}                & 22,617                                    & \multicolumn{1}{c|}{15,000}                                    \\
\multicolumn{1}{r|}{\textit{RITW}}                  & 19,963                                  & \multicolumn{1}{c|}{11,816}                                 \\ \cline{2-3}

\end{tabular}%

}
\vspace{-.5em}
\end{table}

\subsubsection{Experiments}
We performed four experiments:
\begin{description}[leftmargin=.5cm]
    \item[EXP2a] $\mathcal{R}$: A baseline comparison between existing solutions (passive) and our solution (active) in detecting RT-DFs.
    \item[EXP2b] $\mathcal{C}$: An evaluation of the task detection model which ensures that the caller indeed performed the challenge. 
     \item[EXP2c] $\mathcal{I}$: An evaluation of the identity model which ensures that the caller didn't just turn off the RT-DF for the challenge.
    \item[EXP2d] $\mathcal{R,C,I}$: An evaluation of the system end-to-end to evaluate the performance of the system as a whole.
\end{description}
We do not evaluate $\mathcal{T}$ because it is just a restriction that the first frame of the response $r_c$ be received within approximately one second from the start time of the challenge.

To measure the performance of the models, we use the area under the curve (AUC) and equal error rate (EER) metrics. AUC measures the general trade-off between the true positive rate (TPR) and the false positive rate (FPR). An AUC of 1.0 indicates a perfect classifier while an AUC of 0.5 indicates random guessing. The EER captures the trade-off between the FPR and the false negate rate (FNR). A lower EER is better.

\subsection{Experiment Results}

\subsubsection{EXP2a ($\mathcal{R}$)}
The goal of \textbf{EXP2a} was to see if our system can improve the detection of RT-DFs if the adversary is forced to perform a task that is outside of the deepfake model's capabilities. In Table \ref{tab:eval}, we compare the performance of the five deepfake detectors on (1) detecting regular deepfake speech (baseline) and on (2) detecting deepfake challenges. The bold values indicate challenges which improved the performance of the corresponding model. We see that with the exception of SpecRNet, all of the detectors benefit from examining challenges. Overall, the best performing model was GMM-ASVspoof \textit{with} the challenges. This means that the challenges provide a better way to detect RT-DFs.

\begin{table*}[]
\caption{The AUC and EER of deepfake detectors when used as regular deepfake detectors (baseline) and when used as $\mathcal{R}$ with the challenges.}
\label{tab:eval}
\resizebox{0.8\textwidth}{!}{%
\begin{tabular}{lcccccccccc}
\def\arraystretch{.7}

{AUC}                   & \multicolumn{1}{|c|}{Baseline} & R                    & T\&C                 & SE                   & P                    & VS                   & V                    & S                    & HT                   & \multicolumn{1}{c|}{Co}             \\ \hline\hline
\multicolumn{1}{r|}{\textit{SpecRNet}}     & \multicolumn{1}{c|}{0.952}    & 0.914                & 0.538                & 0.796                & 0.825                & 0.922                & 0.92                 & 0.834                & 0.701                & \multicolumn{1}{c|}{0.789}          \\
\multicolumn{1}{r|}{\textit{One-Class}}    & \multicolumn{1}{c|}{0.939}    & \textbf{0.952}       & \textbf{0.967}       & \textbf{0.941}       & \textbf{0.954}       & \textbf{0.958}       & \textbf{0.957}       & \textbf{0.948}       & 0.896                & \multicolumn{1}{c|}{0.832}          \\
\multicolumn{1}{r|}{\textit{GMM-AsvSpoof}} & \multicolumn{1}{c|}{0.949}    & \textbf{0.951}       & \textbf{0.978}       & \textbf{0.953}       & \textbf{0.97}        & \textbf{0.957}       & \textbf{0.949}       & 0.928                & \textbf{0.949}       & \multicolumn{1}{c|}{0.833}          \\
\multicolumn{1}{r|}{\textit{PC-DARTS}}     & \multicolumn{1}{c|}{0.551}    & \textbf{0.568}       & \textbf{0.557}       & \textbf{0.611}       & 0.507                & \textbf{0.586}       & \textbf{0.579}       & \textbf{0.655}       & \textbf{0.675}       & \multicolumn{1}{c|}{\textbf{0.635}} \\
\multicolumn{1}{r|}{\textit{LOF}}          & \multicolumn{1}{c|}{0.678}    & 0.614                & \textbf{0.93}        & 0.635                & \textbf{0.756}       & \textbf{0.771}       & \textbf{0.824}       & 0.593                & \textbf{0.681}       & \multicolumn{1}{c|}{\textbf{0.982}} \\ \hline\hline
                                           & \multicolumn{1}{l}{}          & \multicolumn{1}{l}{} & \multicolumn{1}{l}{} & \multicolumn{1}{l}{} & \multicolumn{1}{l}{} & \multicolumn{1}{l}{} & \multicolumn{1}{l}{} & \multicolumn{1}{l}{} & \multicolumn{1}{l}{} & \multicolumn{1}{l}{}                \\
{EER}                   & \multicolumn{1}{|c|}{Baseline} & R                    & T\&C                 & SE                   & P                    & VS                   & V                    & S                    & HT                   & \multicolumn{1}{c|}{Co}             \\ \hline\hline
\multicolumn{1}{r|}{\textit{SpecRNet}}     & \multicolumn{1}{c|}{0.116}    & 0.163                & 0.475                & 0.285                & 0.261                & 0.155                & 0.154                & 0.245                & 0.354                & \multicolumn{1}{c|}{0.281}          \\
\multicolumn{1}{r|}{\textit{One-Class}}    & \multicolumn{1}{c|}{0.128}    & \textbf{0.123}       & \textbf{0.099}       & 0.133                & \textbf{0.118}       & \textbf{0.112}       & \textbf{0.104}       & \textbf{0.128}       & 0.187                & \multicolumn{1}{c|}{0.259}          \\
\multicolumn{1}{r|}{\textit{GMM-AsvSpoof}} & \multicolumn{1}{c|}{0.122}    & \textbf{0.1}         & \textbf{0.071}       & \textbf{0.099}       & \textbf{0.09}        & \textbf{0.092}       & \textbf{0.115}       & 0.143                & 0.131                & \multicolumn{1}{c|}{0.255}          \\
\multicolumn{1}{r|}{\textit{PC-DARTS}}     & \multicolumn{1}{c|}{0.449}    & \textbf{0.418}       & 0.494                & \textbf{0.386}       & 0.494                & \textbf{0.43}        & \textbf{0.437}       & \textbf{0.366}       & \textbf{0.334}       & \multicolumn{1}{c|}{\textbf{0.415}} \\
\multicolumn{1}{r|}{LOF}                   & \multicolumn{1}{c|}{0.326}    & 0.419                & \textbf{0.122}       & 0.412                & \textbf{0.262}       & \textbf{0.301}       & \textbf{0.26}        & 0.38                 & 0.382                & \multicolumn{1}{c|}{\textbf{0.051}} \\ \hline\hline
\end{tabular}%
}
\end{table*}

\subsubsection{EXP2b ($\mathcal{C}$)}
\begin{figure}
    \centering
    \includegraphics[width=\columnwidth]{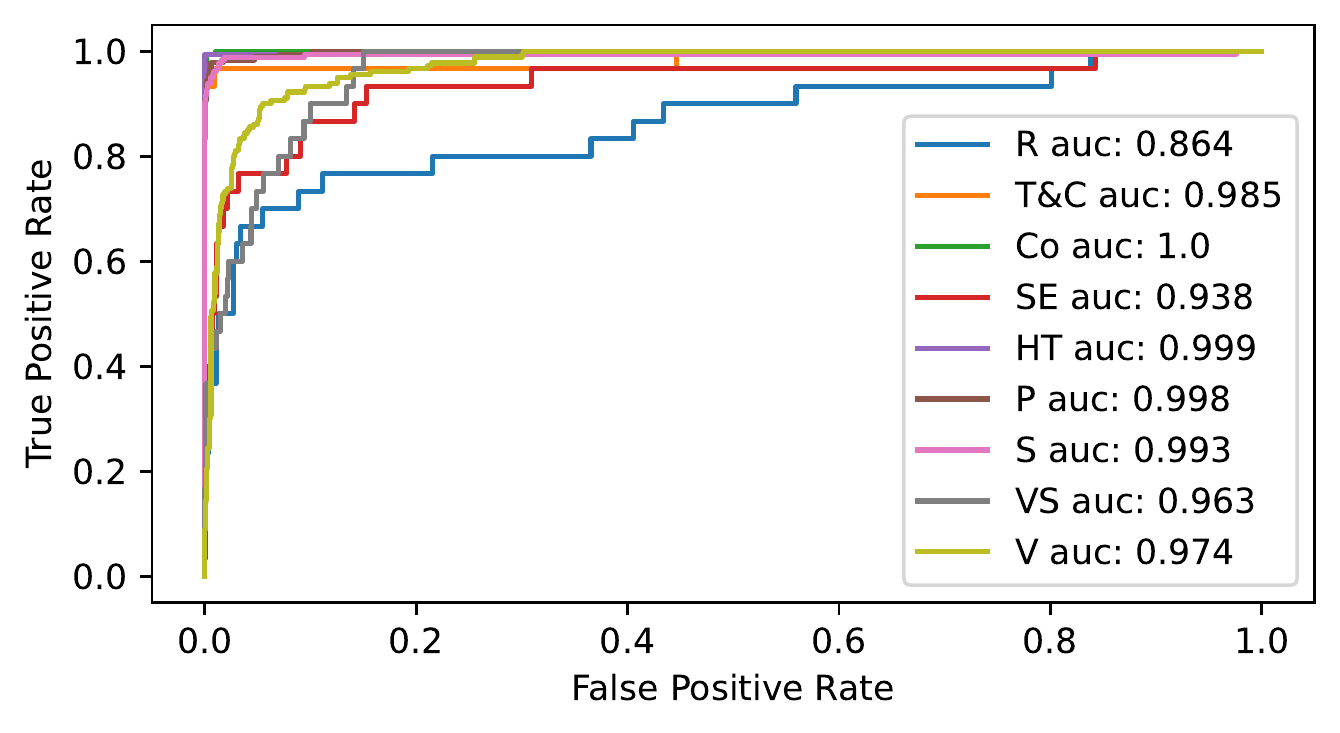}
    \caption{The performance of the task detection model $\mathcal{C}$.}
    \label{fig:exp2b}
\end{figure}

If an attacker is evasive, he may try to do nothing instead of the challenge. It's also possible that the attacker will try the challenge, but the model will output nothing because it can't generate the data. Fig. \ref{fig:exp2b} shows that either way, the task detector $\mathcal{C}$ can tell whether the task was performed or not with high certainty.

\subsubsection{EXP2c ($\mathcal{I}$)}
\begin{figure}
    \centering
    \includegraphics[width=\columnwidth]{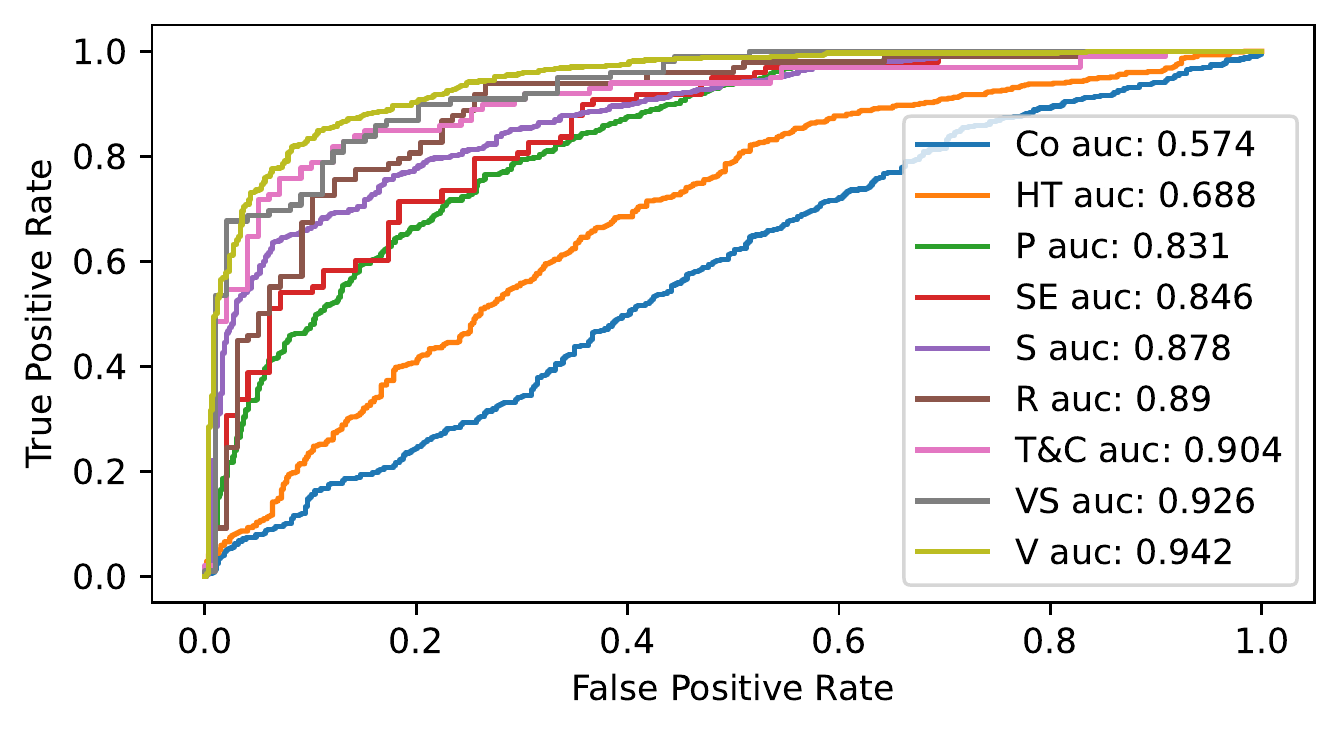}
    \caption{The performance of the unsupervised identity detection model $\mathcal{I}$ for different tasks.}
    \label{fig:exp2c}
\end{figure}

Another evasive strategy is where the attacker turns off the RT-DF while performing the challenge. In this scenario, we compare the identity of the caller before ($a_t$) and during ($r_c$) the challenge. In Fig. \ref{fig:exp2c} we present the results of the identity detector $\mathcal{I}$. Here we can see that the model does quite well, with the exception of the tasks `hum' and `cough' which do not carry much of the speaker's identity.

\subsubsection{EXP2d ($\mathcal{R,I,C}$): D-CAPTCHA}
\begin{figure}
    \centering
    \includegraphics[width=\columnwidth]{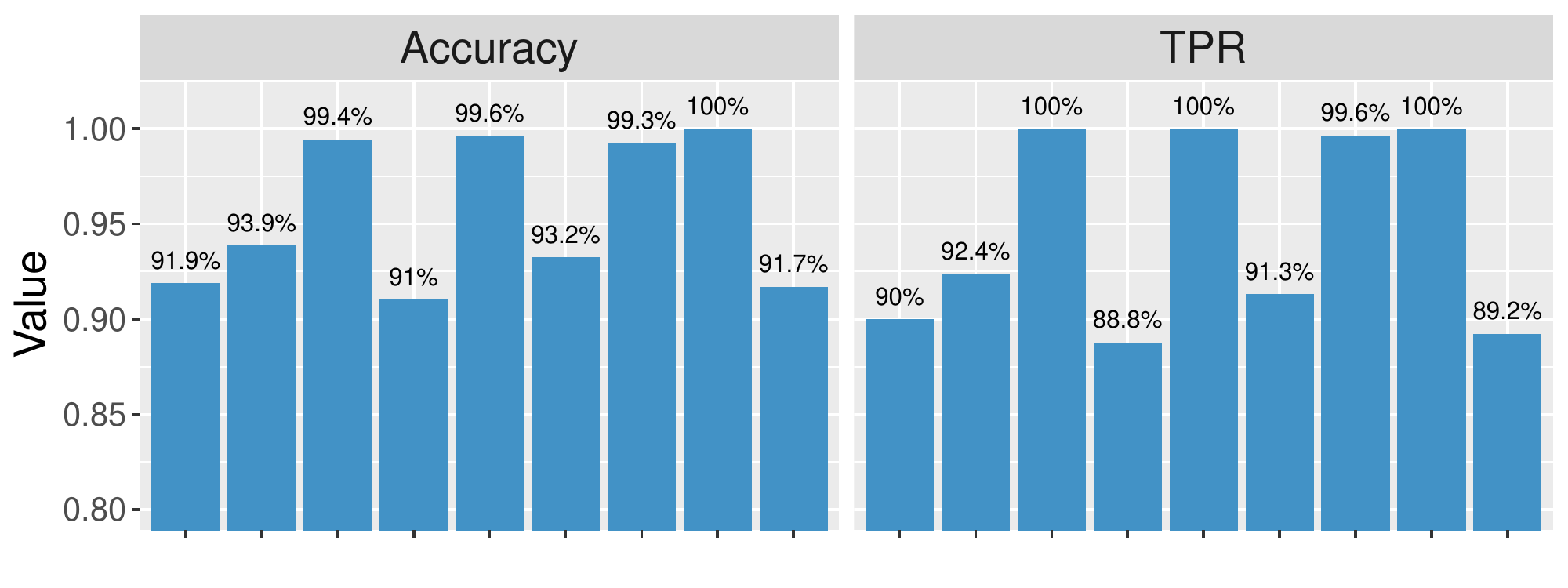}
    \includegraphics[width=\columnwidth]{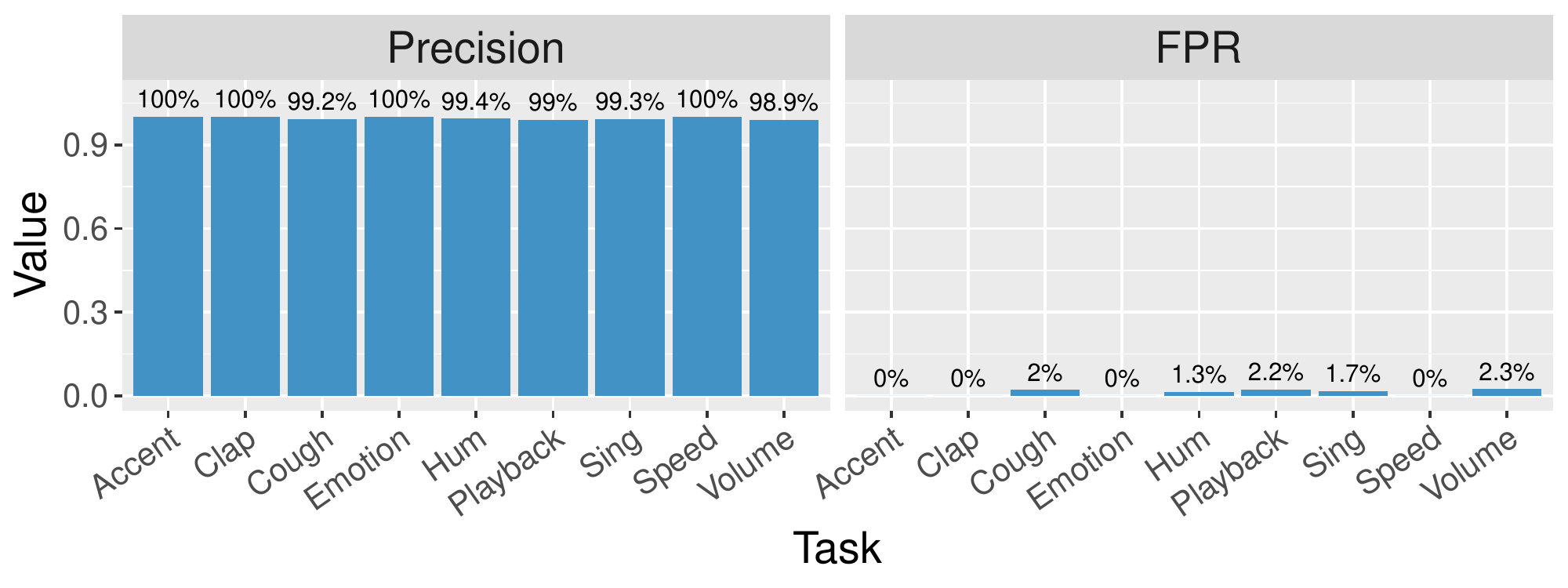}
    \caption{The performance of the ensure D-CAPTCHA system (end-to-end).}
    \label{fig:exp2d}
\end{figure}

Finally, when executing all three models, we must consider how the successes and failures of each model compound together. We set the threshold for each model ($\mathcal{R,I,C}$) so that the FPR=0.01. We then passed through 3,317 real responses and 8,758 deepfake responses. Fig. \ref{fig:exp2d} presents the results. We found that we were able to achieve a TPR of 0.89-1.00. FPR of 0.0-2.3 and accuracy 
of 91-100\% depending on the selected task. In contrast, the model which performed the best on deepfake speech detection (baseline) was SpecRNet with a TPR of 0.66 and accuracy of 71\% when the FPR=0.01. Therefore, D-CAPTCHA significantly outperforms the baseline and provides a good defense against RT-DFs audio calls.

\section{Future work: Video D-CAPTCHA}\label{sec:futurework} 
\begin{figure*}
    \centering
    \includegraphics[width=\textwidth]{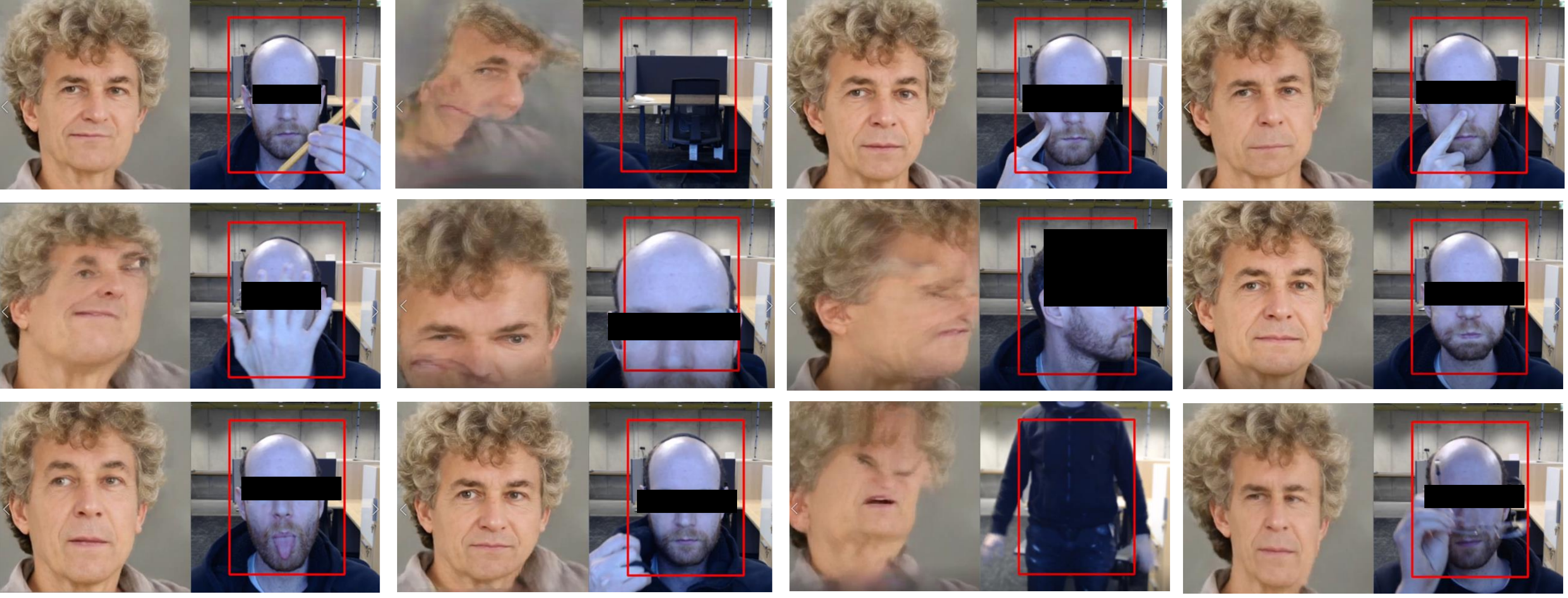}
    \caption{Preliminary results showing how the D-CAPTCHA system can help prevent RT-DF video calls. Here a zero-shot reenactment model called Avatarify breaks the moment the caller performs an action other than basic expressions and talking.}
    \label{fig:videocaptcha}
\end{figure*}
As mentioned in the introduction, the same D-CAPTCHA system outlined in this paper can be applied to video-based RT-DFs as well. For example, to prevent imposters from joining online meetings (such as the cases in \cite{European87:online,Binancee98:online}) we can forward suspicious calls to a D-CAPTCHA system. There are a wide variety of tasks which existing models and pipelines cannot handle for similar reasons to those listed in section \ref{subsec:limitations}. For example, the caller can be asked to drop/bounce objects, fold shirt, stroke hair, interact with background, spill water, pick up objects, perform hand expressions, press on face, remove glasses, turn around, and so on. These tasks can easily be turned into challenges to detect video-based RT-DFs.

To demonstrate the potential, we have performed some initial experiments and will now present some preliminary results. In our experiment we used a popular zero-shot RT-DF model called Avatarify\footnote{\url{https://github.com/alievk/avatarify-python}} based on the work of \cite{NIPS2019_8935} to reenact (puppet) a single photo. We were able to achieve a realistic RT-DF video at 35 fps with negligible distortions if the face stayed in a frontal position. However, when we performed some of the mentioned challenges, the model failed and large distortions appeared. Fig., \ref{fig:videocaptcha} in the appendix presents some screenshots of the video during the challenges.

These preliminary results indicate that D-CAPTCHAs can be a good solution for both RT-DF audio and video calls.


\section{Conclusion}
Deepfakes are rapidly improving in terms of quality and speed. This poses a significant threat as attackers are already using real-time deepfakes to impersonate people over calls. Current defenses use passive methods to identify deepfakes via their flaws. However, this approach may have limits as the quality of deepfakes continues to advance. Instead, in this work we proposed an active defense strategy: D-CAPTCHA. By challenging the attacker to create content under four constraints based on practical and technological limitations, we can force the deepfake model to expose itself. By protecting calls and meetings from deepfake imposters, we believe that this system can significantly improve the security of organizations and individuals.

\begin{acks}
This work was supported by the U.S.-Israel Energy Center managed by the Israel-U.S. Binational Industrial
Research and Development (BIRD) Foundation and the Zuckerman STEM Leadership Program.
\end{acks}

\bibliographystyle{ACM-Reference-Format}
\bibliography{paper.bib}


\appendix

\section{Ethical Disclosures}
The experiments performed in this study have received our institution's ethical committee's approval. All 20 volunteers whose voices were used to create deepfakes permitted the use of their data for this purpose. To protect our volunteers, the trained RT-DF voice models will not be shared. 

\section{Additional Figures}
\begin{figure}[h]
    \centering
    \includegraphics[width=.6\columnwidth]{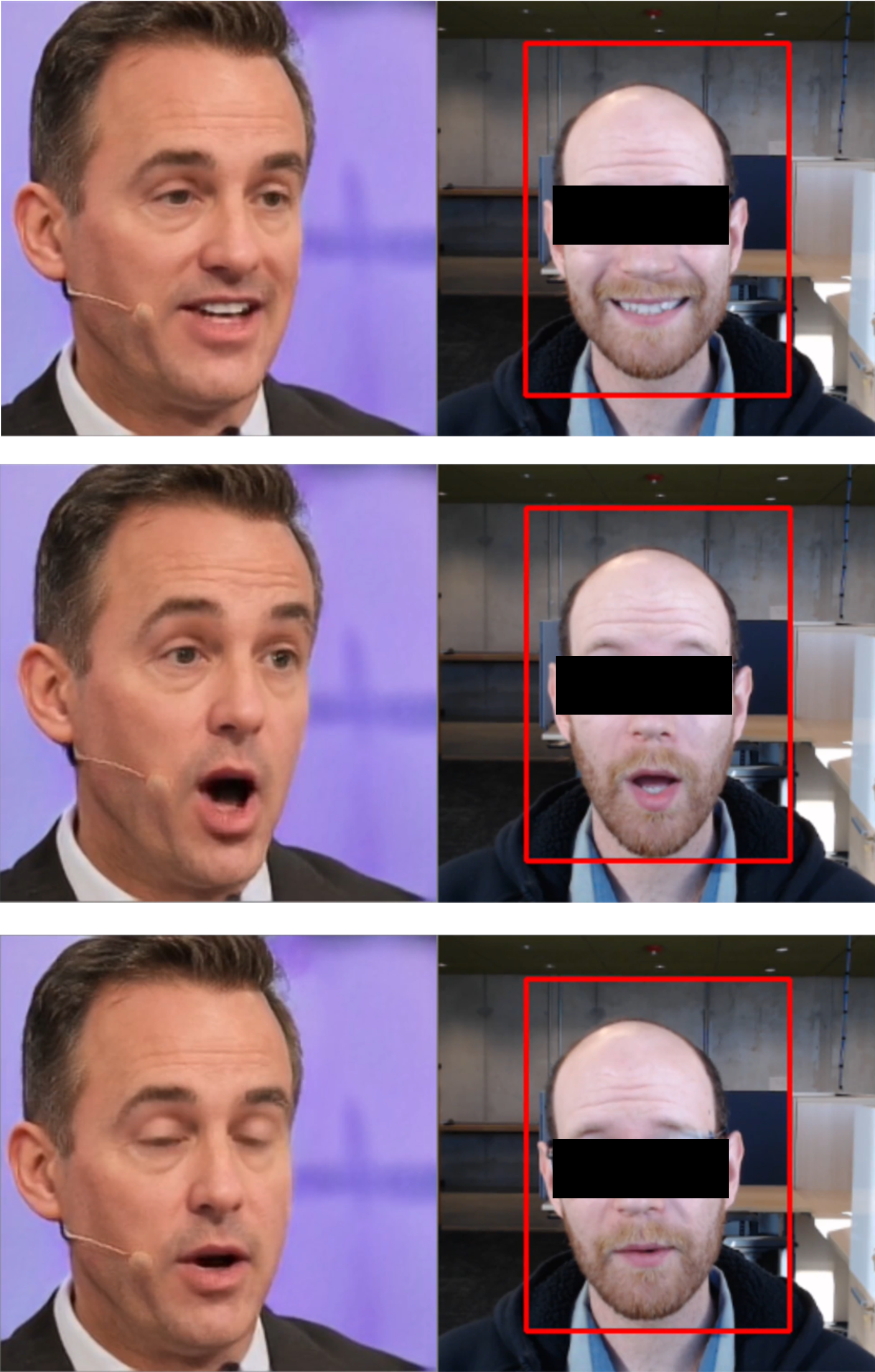}
    \caption{Examples of the Avatarify zero-shot RT-DF model working as expected. Here there are no significant anomalies because the caller has a frontal pose and is talking normally.}
    \label{fig:ceo}
\end{figure}

\begin{figure*}
    \centering
    \includegraphics[width=.8\columnwidth]{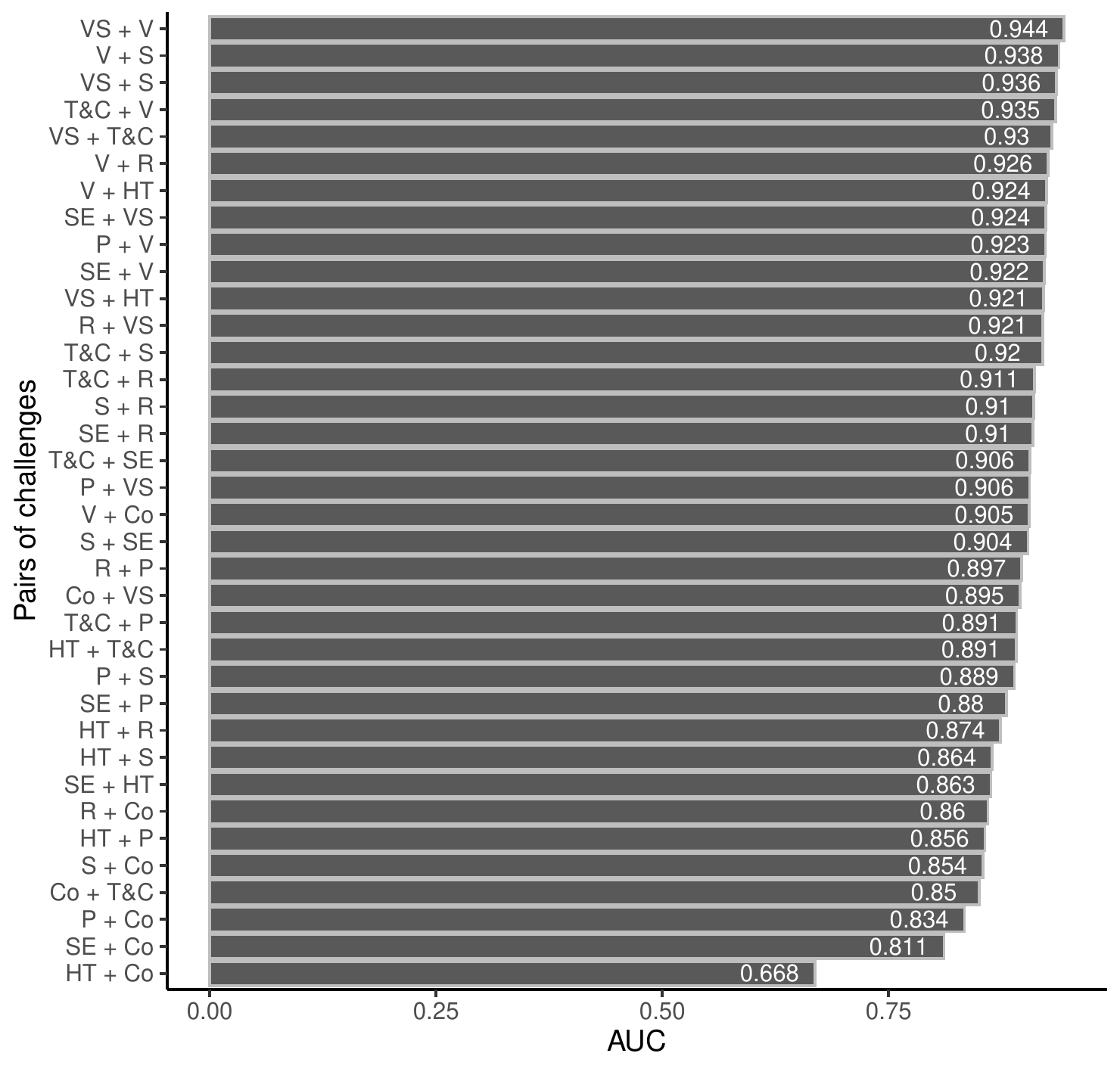}
    \caption{The performance of $\mathcal{I}$ when two challenges are requested, measured in AUC.}
    \label{fig:identity2}
\end{figure*}

\begin{figure*}[]
        \centering
    \includegraphics[width=.75\textwidth]{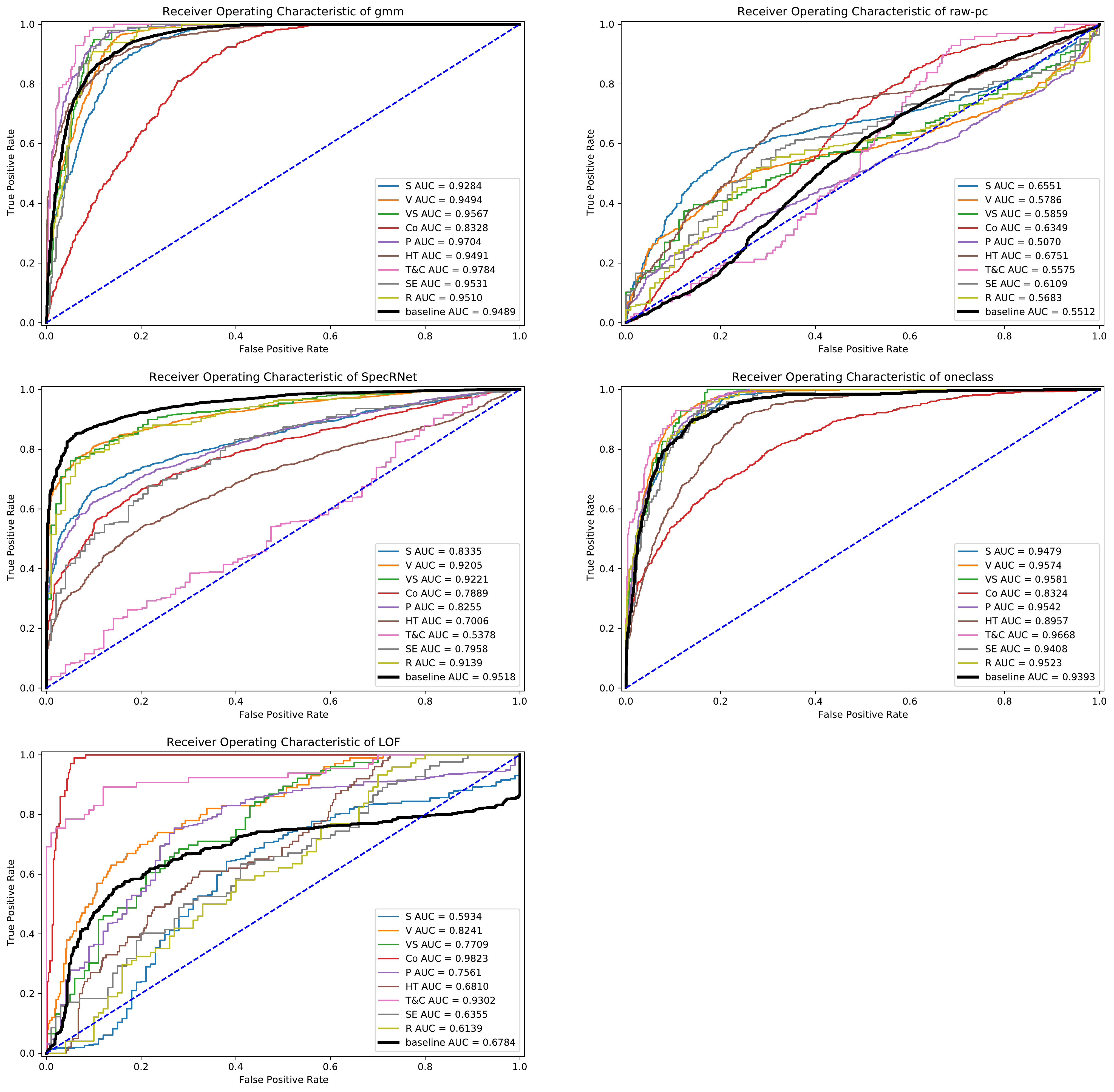}
    \caption{ROC plots for each deepfake detection model from experiment \textbf{EXP2a}. The bold line shows the baseline (regular deepfake detection) and the others show the performance on the given task.}
    \label{fig:rocr_baselines}
\end{figure*}

\end{document}